\documentclass[11pt,english,a4paper,floatfix]{article}
\pdfoutput=1
\usepackage{jcappub}

\usepackage{caption}
\usepackage{bm}
\usepackage{amsfonts}
\usepackage{subfig,graphicx}
\usepackage{url}
\usepackage{color}
\usepackage{multirow}
\usepackage{adjustbox}
\usepackage{xpatch}
\usepackage[normalem]{ulem}
\usepackage{todonotes}

\usepackage{array}
\newcommand{\PreserveBackslash}[1]{\let\temp=\\#1\let\\=\temp}
\newcolumntype{C}[1]{>{\PreserveBackslash\centering}p{#1}}

\begin{document}

\title{Sterile neutrino self-interactions: $H_0$ tension and short-baseline anomalies}

\author[a,b]{Maria~Archidiacono}
\affiliation[a]{Universit\'a degli Studi di Milano, via G. Celoria 16, 20133 Milano, Italy}
\affiliation[b]{INFN, Sezione di Milano, Milano, Italy}
\emailAdd{maria.archidiacono@unimi.it}

\author[c,d]{, Stefano~Gariazzo}
\affiliation[c]{Instituto de F{\'\i}sica Corpuscular  (CSIC-Universitat de Val{\`e}ncia), Parc Cient{\'\i}fic UV, C/ Catedr{\'a}tico Jos{\'e} Beltr{\'a}n, 2, E-46980 Paterna (Valencia), Spain}
\affiliation[d]{INFN, Sezione di Torino, Via P. Giuria 1, I--10125 Torino, Italy}
\emailAdd{gariazzo@to.infn.it}

\author[d]{, Carlo~Giunti}
\emailAdd{giunti@to.infn.it}

\author[e]{, Steen~Hannestad}
\affiliation[e]{Department of Physics and Astronomy, Aarhus University, DK-8000 Aarhus C, Denmark}
\emailAdd{sth@phys.au.dk}

\author[e]{, Thomas~Tram}
\emailAdd{thomas.tram@phys.au.dk}

\abstract{
Sterile neutrinos with a mass in the eV range have been invoked as a possible explanation of a variety of short baseline (SBL) neutrino oscillation anomalies.
However, if one considers neutrino oscillations between active and sterile neutrinos,
such neutrinos would have been fully thermalised in the early universe,
and would be therefore in strong conflict with cosmological bounds.
In this study we first update cosmological bounds on the mass and energy density of eV-scale sterile neutrinos.
We then perform an updated study of a previously proposed model in which the sterile neutrino couples to a new light pseudoscalar degree of freedom.
Consistently with previous analyses, we find that the model provides a good fit to all cosmological data and allows the high value of $H_0$ measured in the local universe to be consistent with measurements of the cosmic microwave background.
However, new high $\ell$ polarisation data constrain the sterile neutrino mass to be less than approximately 1~eV in this scenario.
Finally, we combine the cosmological bounds on the pseudoscalar model with a Bayesian inference analysis of SBL data and conclude that only a sterile mass in narrow ranges around 1~eV remains consistent with both cosmology and SBL data.
}

\maketitle

\section{Introduction}
\label{sec:into}

Recent cosmological results have confirmed the success of the Lambda Cold Dark Matter ($\Lambda$CDM) model over a wide range of redshifts and scales.
Nevertheless, as the precision of cosmological data increases, tensions among different data sets appear, most notably the $4.4\,\sigma$ tensions between the value of the Hubble constant $H_0$ in the early Universe and the one in the late Universe. The former is inferred from the Cosmic Microwave Background (CMB) data by Planck \cite{Aghanim:2018eyx}. The latter comes from direct measurement of various observables with different techniques\footnote{See Ref.\ \cite{Riess:2020sih,Verde:2019ivm} for recent reviews on the $H_0$ tension.}: 
building a cosmic distance ladder by calibrating the SuperNovae luminosity distance with variable stars (e.g., the Cepheids of the SHOES project \cite{Riess:2019cxk}, and the Mira variables \cite{Huang:2019yhh}) or with the Tip of the Red Giant Branch \cite{Yuan:2019npk}, using gravitational waves from neutron star mergers as standard sirens \cite{Abbott:2017xzu}, or the different ages of galaxies as cosmic clocks \cite{Jimenez:2001gg,Moresco:2016mzx}, estimating the time delay distance between multiple images of distant quasars induced by strong gravitational lensing (H0LiCOW \cite{Wong:2019kwg}), measuring the angular diameter distance by observing water maser in an accretion disk of supermassive black holes \cite{Pesce:2020xfe},
or using the baryonic Tully-Fisher relation \cite{Schombert:2020pxm}.
This tension has generated a strong debate whether it is caused by unaccounted systematics \cite{Bernal:2018cxc} or it is the signal of new physics\footnote{For a discussion of possible solutions see \cite{Knox:2019rjx} and references therein.} either in the early Universe \cite{Smith:2019ihp} or in the late Universe, with the latter option being less viable due to the SN and Baryonic Acoustic Oscillations (BAO) constraints at $z \lesssim 2$ \cite{Lemos:2018smw,Benevento:2020fev,Dhawan:2020xmp}.

One possibility which modifies both the early- and late-time Universe is the addition of light sterile neutrinos.
Such particles have been hinted at by a number of terrestrial experiments.
The effect of such neutrinos on cosmological observables has been investigated numerous times in the literature (see e.g.\ 
\cite{Hamann:2010bk,Motohashi:2012wc,Giusarma:2011zq,Hamann:2011ge,Giusarma:2011ex,Hagstotz:2020ukm,Feng:2019jqa,Vegetti:2018dly,Zhao:2017urm,Tang:2014yla,Vincent:2014rja,Archidiacono:2013xxa,Joudaki:2012uk,Archidiacono:2012ri}).
With the advent of precision CMB measurements by the Planck mission it became clear that light sterile neutrinos in themselves do not resolve the $H_0$ tension.
However, we previously demonstrated that sterile neutrinos with a coupling to a new, massless pseudoscalar alleviate the $H_0$ tension and lead to a significantly improved fit to observables \cite{Archidiacono:2014nda,Archidiacono:2015oma,Archidiacono:2016kkh}%
\footnote{
Neutrino interactions either of a different kind (vector-like sterile neutrino self-interactions \cite{Hannestad:2013ana,Dasgupta:2013zpn}) or extended to the active neutrino sector \cite{Archidiacono:2013dua,Forastieri:2019cuf,Kreisch:2019yzn,Park:2019ibn,Escudero:2019gvw,Berbig:2020wve,Blinov:2020hmc} have been extensively studied over the past few years.
However, most of these models turns out to be severely constrained either by cosmological bounds on the sum of neutrino masses and on the free-streaming of active neutrinos \cite{Mirizzi:2014ama,Saviano:2014esa,Chu:2015ipa,Song:2018zyl,Chu:2018gxk,Grohs:2020xxd}, or by laboratory limits \cite{Arcadi:2018xdd,Blinov:2019gcj}.
}.

Here we re-examine the effect of sterile neutrinos, both with and without additional self-interactions, on cosmology.
The aim is to provide updated constraints on our Pseudoscalar model in light of the latest Planck measurements, and in combination with recent results from ground-based sterile neutrino searches. To this end we proceed as follows:
Section \ref{sec:model} contains an outline of the model framework we use.
Section \ref{sec:methodology} provides a description of our parameter estimation methodology.
In Section \ref{sec:results} we present our main cosmological results, and in Section \ref{sec:SBL} we combine them with an up-to-date global fit of neutrino oscillation experiments.
Finally in Section \ref{sec:conclusions} we draw our conclusions.

\section{Pseudoscalar sterile neutrino interactions}
\label{sec:model}

The models we investigate all have a light, sterile neutrino in addition to the standard neutrino sector.
Besides providing updated constraints on the presence of such particles from the latest cosmological data, we will also reinvestigate constraints on a model in which the sterile neutrinos are coupled to a new, very light pseudoscalar degree of freedom.

\subsection{``Vanilla'' sterile neutrinos}

We take the simple model which has a non-coupled sterile neutrino to be described in terms of only two parameters: the physical sterile neutrino mass, $m_s$%
\footnote{
A sterile neutrino is a flavor eigenstate and it has no definite mass.
Here we approximate $m_4\simeq m_s$ thanks to the fact that, in 3+1 models,
the fourth mass eigenstate is mainly mixed with the sterile flavour eigenstate.
},
and the effective contribution to the energy density in the relativistic regime, $\Delta N_{\rm eff}$.
While this parametrisation does not completely capture all physical aspects of the sterile sector (for example the exact relation between $\Delta N_{\rm eff}$, $m_s$, and the late-time contribution to the energy density $\Omega$ depends on the specific late-time sterile distribution function), it is adequate for the analysis of current cosmological data. We will assume that the sterile neutrino has the same temperature as standard model neutrinos%
\footnote{
This assumption is corroborated by results obtained e.g.\ in \cite{Gariazzo:2019gyi}.
}
so that the energy density in the relativistic regime is given by
\begin{equation}
\rho_{\rm s} = \Delta N_{\rm eff} \rho_{\nu},
\end{equation}
and in the non-relativistic regime by
\begin{equation}
\rho_s = \Delta N_{\rm eff} m_s n_{\nu},
\end{equation}
where $\rho_\nu$ and $n_\nu$ are the energy density and number density of completely decoupled standard model neutrinos.

\subsection{Pseudoscalar interactions}
\label{subs:pse_model}

Next, we will derive current constraints on a model first proposed in \cite{Archidiacono:2014nda}.
In this model the sterile neutrinos couple to a new, and effectively massless pseudoscalar degree of freedom.
Such an additional interaction leads to rapid pair-annihilation and disappearance of the sterile neutrinos below a temperature corresponding to the mass.
This was first suggested as a way of reconciling a large mass in the active neutrino sector with cosmological measurements, a model dubbed the ``neutrinoless universe'' \cite{Beacom:2004yd}.
However, while this model does avoid an overly large suppression of the matter power spectrum, it leads to very significant changes to the CMB spectrum, and at present it is unclear whether the model can be reconciled with cosmological data.

Contrary to this, it was shown in \cite{Archidiacono:2014nda} that if the pseudoscalar couples only to the sterile neutrino a very good fit to cosmological data can be obtained.
Furthermore, it was demonstrated that the model predicts a value of the Hubble parameter almost exactly identical to the one measured in the local universe.
This finding was confirmed in \cite{Archidiacono:2016kkh,Archidiacono:2015oma} which considered more recent data sets.

Our parameterisation of the model is as follows (see \cite{Archidiacono:2015oma} for a more detailed discussion): 

\begin{itemize}

\item Around the epoch of standard model neutrino freeze-out the total energy density in standard model neutrinos, sterile neutrinos, and pseudoscalars is given by $N_{\rm eff}$, where $N^{\rm std}_{\rm eff}=3.046$~\cite{Mangano:2005cc} (see also \cite{deSalas:2016ztq,Gariazzo:2019gyi,Bennett:2019ewm,Akita:2020szl}) for pure standard model neutrinos. However, we will in general allow for extra energy density in the combined fluid so that $N_{\rm eff}$ can be larger.
As demonstrated in \cite{Archidiacono:2014nda}, if the dimensionless coupling is larger than $g \sim 10^{-6}$, the production of sterile neutrinos is delayed until the time of active neutrino decoupling, which roughly coincides with the onset of Big Bang Nucleosyntheisis. Therefore, the BBN bounds on $N_{\rm eff}$ are evaded \cite{Schoneberg:2019wmt}.

\item Subsequent to neutrinos decoupling from the electromagnetically interacting plasma, the energy in the neutrino-pseudoscalar sector is redistributed by oscillations, so that the sterile plus pseudoscalar sector ends up with a fraction of $(4/7+1)/(4/4+4)=11/32$ of the total energy density, while the remaining fraction 21/32 goes in the active sector.

\item After this happens, the active and the sterile-pseudoscalar components are completely separated and do not exchange neither energy nor momentum,
provided that the dimensionless coupling is larger than $g \sim 10^{-6}$.
In this case the sterile neutrinos and pseudoscalars become very strongly coupled prior to the sterile neutrinos becoming non-relativistic. Therefore, the combined system can be treated as a single fluid with a well-defined energy density and equation of state.
Once the temperature drops below the sterile neutrino mass, its entropy is transferred to the pseudoscalar so that any sterile rest mass is converted to additional energy in the pseudoscalar component.

\end{itemize}

\section{Methodology and data}
\label{sec:methodology}

We compute the theoretical predictions for the cosmological quantities by means
of the Boltzmann solver
\texttt{CLASS}~\cite{Lesgourgues:2011re,Blas:2011rf,Lesgourgues:2011rh},
while its python counterpart
\texttt{MontePython}~\cite{Audren:2012wb,Brinckmann:2018cvx}
is responsible for computing the cosmological likelihoods
and performing the Markov Chain Monte Carlo (MCMC).
In order to consider the model described in subsection~\ref{subs:pse_model},
we modified the public \texttt{CLASS} code to take into account the presence of the new particle.

We compute the constraints with the MCMC generator provided by \texttt{MontePython}
and scan the parameter space of our cosmological models.
Being both models based on the standard $\Lambda$CDM model,
they have six free parameters in common:
the baryon and cold dark matter energy densities ($\omega_{\rm b}$, $\omega_{\rm c}$),
the angular size of the sound horizon at recombination ($\theta_s$),
the reionization optical depth ($\tau$),
the amplitude and tilt of the spectrum of primordial curvature fluctuations ($A_{\rm s}$, $n_{\rm s}$).
All these parameters are sampled with a uniform prior without bounds.
The additional parameters used to describe the neutrino sector are
the mass of the sterile neutrino $m_s$,
and the number of additional relativistic degrees of freedom $\Delta N_{\rm eff}$.

Within this extended ensemble of parameters we explore three different neutrino scenarios:
\begin{description}
\item[Vanilla:]
the active neutrino sector is described by massless neutrinos providing the standard $N^{\rm std}_{\rm eff}=3.046$~\cite{Mangano:2005cc}, which is kept fixed; the effect of sterile neutrinos is embedded into $\Delta N_{\rm eff}\geq0$, which is free to vary, as well as its mass $0\leq m_s/{\rm eV}\leq 10$;
\item[Thermal:] same as Vanilla, but here $\Delta N_{\rm eff}$ is fixed to $1$ to reproduce the case of one fully thermalized sterile neutrino;
\item[Pseudoscalar:] (sometimes abbreviated to ``Pseudo'',): as in the Vanilla case, we have
a total $N_{\rm eff}=N^{\rm std}_{\rm eff}+\Delta N_{\rm eff}$ given by the fixed $N^{\rm std}_{\rm eff}=3.046$ plus the varying $\Delta N_{\rm eff}\geq0$, however here the total $N_{\rm eff}=N^{\rm std}_{\rm eff}+\Delta N_{\rm eff}$ energy density is split into a fraction $21/32$ for the active (massless) sector and a fraction $11/32$ that represents the pseudoscalar - sterile fluid. The sterile neutrino has again a mass $0\leq m_s/{\rm eV}\leq 10$ (see Section \ref{sec:model}).
\end{description}

The constraints on our parameters are computed by fitting CMB and $H_0$ observations.
We use the most recent measurements of the CMB temperature, polarisation, and lensing spectra by
Planck \cite{Akrami:2018vks,Aghanim:2018eyx}.
The computational details on the Planck likelihood are extensively reported in \cite{Aghanim:2019ame}.
In order to disentangle the effects of temperature and polarization, we consider either ``Planck TT'', which includes low multipoles information on temperature and polarisation and only temperature measurements at high multipoles,
or ``Planck TTTEEE'', which also includes polarisation at high multipoles.
We also test the impact of a prior that takes into account the most recent determination of the Hubble parameter today,
$H_0 = 74.03 \pm 1.42$~km/s/Mpc from \cite{Riess:2019cxk} (hereafter we will refer to this prior as R19).
Finally, we add ``Planck lensing'' and a combination of Baryonic Acoustic Oscillations (BAO)\footnote{The BAO considered here are: BOSS DR12 \cite{Alam:2016hwk}, and the low redshift (hereafter dubbed ``small z'') 6dFGS \cite{Beutler:2011hx} and SDSS MGS \cite{Ross:2014qpa}.} to test the impact of the information on low redshift matter distribution.

\section{Results}
\label{sec:results}

In Table \ref{tab:results} we report the mean values with 68\% intervals or the 95\% upper limits for the neutrino parameters, for $H_0$ and for $n_s$ obtained by fitting five data combinations (Planck TT-only, Planck TTTEEE, Planck TTTEEE + R19, Planck TTTEEE + lensing + BAO, and Planck TTTEEE + lensing + BAO + R19) to the three models described above.

\begin{table}[tbp]
\begin{adjustbox}{width=1.\textwidth}
\centering
 \begin{tabular}{| l | c | c |  c ||  c | c | c |}
 \hline
& Vanilla & Pseudo & Thermal & Vanilla & Pseudo & Thermal\\ 
 \hline
 Parameter &\multicolumn{3}{ c ||}{Planck TT}\\
 \cline{1-4}
  \rule{0pt}{1.1\normalbaselineskip}$\Delta N_{\rm eff}$& $<0.28$ & $<0.86$ & $1$ \\[1em]
 $m_s\,[{\rm eV}]$& $<8.77$ & $3.1_{-1.1}^{+1.3}$ & $<0.44$ \\[1em]
 $H_0\,[{\rm km}/{\rm s}/{\rm Mpc}]$& $67.5_{-1.1}^{+1.0}$ & $72.2_{-2.9}^{+1.7}$ & $74.2_{-1.2}^{+2.1}$ \\[1em]
 $n_s$ & $0.964_{-0.007}^{+0.007}$ & $0.971_{-0.013}^{+0.014}$ & $1.002_{-0.006}^{+0.007}$\\[0.5em]
 \hline 
 Parameter &\multicolumn{3}{ c ||}{Planck TTTEEE}&\multicolumn{3}{ c |}{... + R19}\\
 \hline
 \rule{0pt}{1.1\normalbaselineskip} $\Delta N_{\rm eff}$ & $<0.20$ & $<0.56$ & $1$ & $<0.47$ & $0.38_{-0.15}^{+0.15}$ & $1$\\[1em]
 $m_s\,[{\rm eV}]$ & n.c. & $<1.14$ & $<0.91$ & $<7.58$ & $<1.19$ & $<0.22$\\[1em]
 $H_0\,[{\rm km}/{\rm s}/{\rm Mpc}]$ & $67.8_{-0.7}^{+0.7}$ & $71.6_{-1.6}^{+1.1}$ & $73.3_{-0.5}^{+2.1}$ & $69.6_{-1.3} ^{+0.8}$ & $72.8_{-1.2}^{+1.1}$ & $74.1_{-0.7}^{+0.9}$\\[1em]
 $n_s$ & $0.965_{-0.005}^{+0.005}$ & $0.951_{-0.008}^{+0.006}$ & $0.999_{-0.004}^{+0.007}$ & $0.975_{-0.008}^{+0.006}$ & $0.957_{-0.006}^{+0.006}$ & $1.001_{-0.004}^{+0.004}$\\[0.5em]
 \hline
 Parameter &\multicolumn{3}{ c ||}{... + lensing + BAO}  &\multicolumn{3}{ c |}{... + lensing + BAO + R19}\\
 \hline
  \rule{0pt}{1.1\normalbaselineskip}$\Delta N_{\rm eff}$& $<0.14$ & $<0.41$ & $1$ &$<0.43$ & $0.34_{-0.15}^{+0.14}$ & $1$\\[1em]
 $m_s\,[{\rm eV}]$& n.c. & $<1.03$ & $<0.28$ & n.c. & $<1.08$ & $<$0.24$ $\\[1em]
 $H_0\,[{\rm km}/{\rm s}/{\rm Mpc}]$& $68.1_{-0.5}^{+0.4}$ & $70.0_{-1.1}^{+0.7}$ & $73.3_{-0.7}^{+0.8}$ & $69.2_{-1.1}^{+0.5}$ & $71.4_{-1.0}^{+0.9}$ & $73.5_{-0.6}^{+0.7}$\\[1em]
 $n_s$ & $0.966_{-0.004}^{+0.004}$ & $0.944_{-0.006}^{+0.005}$ & $0.999_{-0.004}^{+0.004}$ & $0.972_{-0.007}^{+0.005}$ & $0.950_{-0.005}^{+0.005}$ & $0.999_{-0.004}^{+0.004}$\\[0.5em]
 \hline
 \hline
\end{tabular}
\end{adjustbox}
\caption{
68\% intervals or 95\% upper limits for neutrino parameters, $H_0$ and $n_s$, obtained by fitting Planck TT-only, Planck TTTEEE, Planck TTTEEE + R19, Planck TTTEEE + lensing + BAO, and Planck TTTEEE + lensing + BAO + R19, to the Vanilla, Pseudoscalar, and Thermal models.
}
\label{tab:results}
\end{table}

The 95\% upper bounds on $\Delta N_{\rm eff}$ are about a factor $3$ tighter in the Vanilla case with respect to the Pseudoscalar model, and in both scenarios the inclusion of high-$\ell$ E-mode polarisation makes them more stringent than when considering Planck TT only.
The bounds are further tightened by adding lensing and BAO, while the $H_0$ prior acts in the opposite direction.
Indeed, as expected, the inclusion of R19 relaxes the limit on $\Delta N_{\rm eff}$ in the Vanilla case, while in the Pseudoscalar model it leads to a $>2\,\sigma$ evidence in favor of a non-zero value of the extra relativistic component.
The reason why the bounds on $\Delta N_{\rm eff}$ are relaxed in the Pseudoscalar model with respect to the Vanilla model is that the extra relativistic component is not free-streaming, hence it does not induce the typical phase shift and suppression of the CMB acoustic peaks \cite{Bashinsky:2003tk} (we will further discuss the impact of free-streaming vs. interacting $\Delta N_{\rm eff}$ on the CMB in Section \ref{sec:phenomenology}).

In the Vanilla case, the physical mass is either unconstrained (Planck TTTEEE) or it has a limit that is very much affected by the prior boundaries, as a consequence of the well known fact that large $m_s$ values are allowed if $\Delta N_{\rm eff}$ is close to zero.
In the Thermal case, the 95\% upper bound on $m_s$ from TTTEEE is a factor $2$ larger than with TT-only, simply because looser constraints help to resolve tensions that are more dramatic in the presence of polarization data; this mechanism is broken by the inclusion of lensing and BAO, and the most stringent bound is obtained with the R19 prior because of the anti-correlation between the hot dark matter density and the Hubble constant. 
In the Pseudoscalar model, there is an evidence for a non zero sterile neutrino mass when fitting only TT: $m_s=3.1_{-1.1}^{+1.3}$~eV (68\% c.l.).
This evidence is removed by the inclusion of the high-$\ell$ E-mode polarisation data, which restricts the range of the sterile neutrino mass values to $m_s<1.14$~eV at 95\% c.l..
The additional R19 prior does not change the result significantly ($m_s<1.19$~eV at 95\% c.l). This point has an impact on the consistency between cosmology and sterile neutrino searches with oscillation experiments, thus, it deserves a dedicated discussion (see next Section \ref{sec:phenomenology}).
It should also be noted here that the constraint on the pseudoscalar mass is much tighter than found in our previous analysis \cite{Archidiacono:2016kkh} due to the addition of the new high-$\ell$ polarisation data.

The slope of the primordial power spectrum is consistent with the $\Lambda$CDM values from Planck 2018 \cite{Aghanim:2018eyx} in the Vanilla case, while the Thermal case prefers a scale-invariant Harrison-Zeldovich power spectrum.
Notice that in the Pseudoscalar fit of TT-only the $n_s$ mean value is slightly shifted towards larger values and, what is most, the $1\,\sigma$ uncertainty is twice larger than in the Vanilla and Thermal cases. This broader range of allowed $n_s$ values shrinks once polarization is taken into account; we will further elaborate on this, in connection with the $m_s$ bounds, in the next Section \ref{sec:phenomenology}.

\begin{figure}[tbp]
\centering 
\includegraphics[width=.45\linewidth]{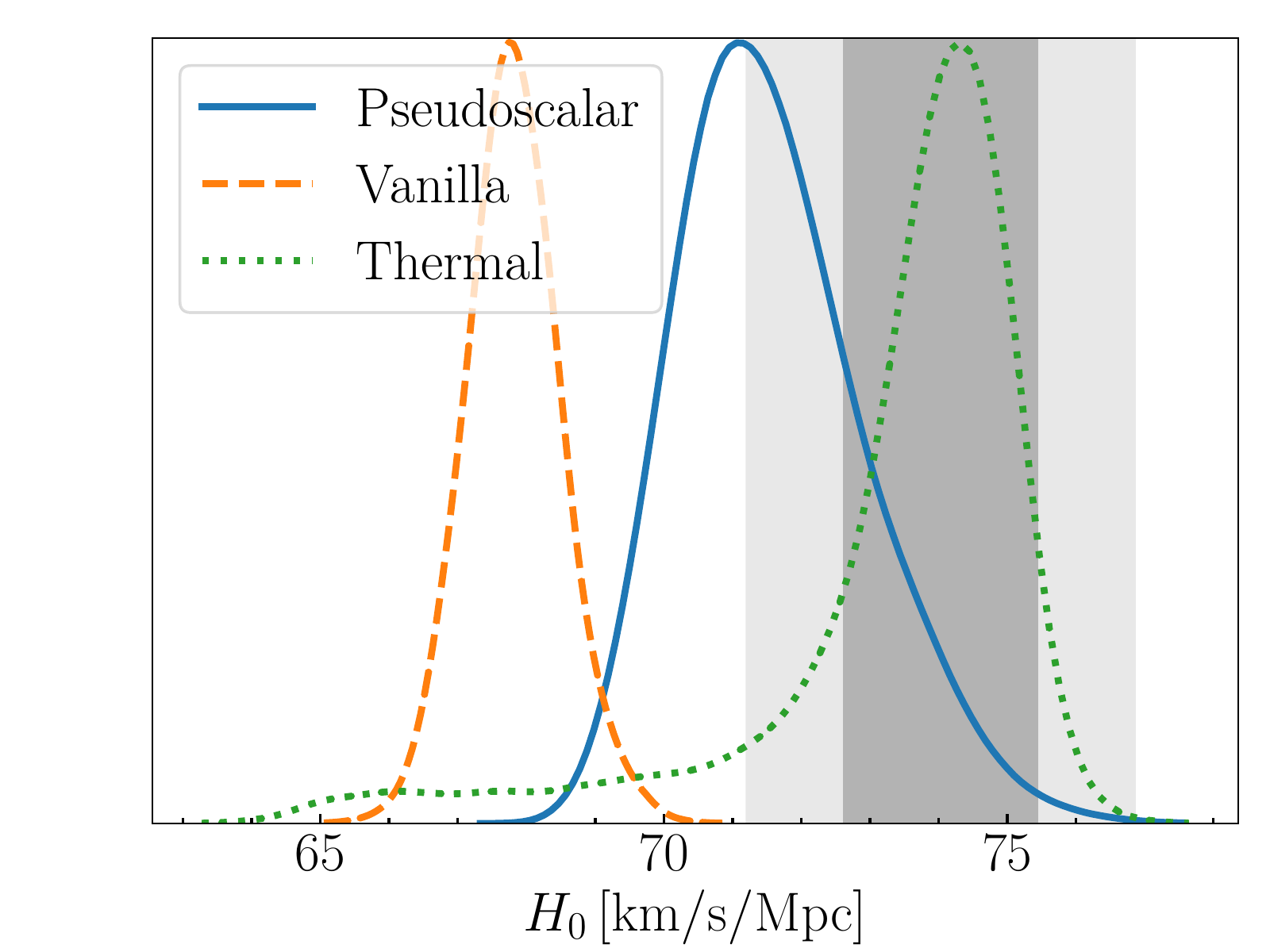}
\caption{\label{fig:H0} Marginalized 1D posterior of $H_0$ obtained by fitting Planck TTTEEE data.
}
\end{figure}
In Figure \ref{fig:H0} we show the unnormalized posterior for $H_0$ obtained by fitting Planck TTTEEE to the three models.
Concerning the mean values of $H_0$, for every data combination we have $H_0^{\rm Vanilla}< H_0^{\rm Pseudoscalar}<H_0^{\rm Thermal}$. The opposite inequality holds for the number of $\sigma$ of tension with R19. Indeed, in the Thermal scenario the tension is always below $1\,\sigma$, while the Vanilla scenario has a tension of $3.8\,\sigma$ ($3.9\,\sigma$) when fitting TT (TTTEEE), which exacerbates when adding lensing and BAO, and which remains significant even including the R19 prior ($2.7\,\sigma$). Concerning our Pseudoscalar model, the tension is below $1\,\sigma$ when fitting TT only, while it becomes slightly larger including polarisation ($1.3\,\sigma$), and it increases above $2\,\sigma$ with lensing and BAO; applying the R19 prior, with and without large scale structures, the tension decreases below $1\,\sigma$.
\begin{table}[tbp]
\begin{adjustbox}{width=1.\textwidth}
\centering
\begin{tabular}{| l | c | c |  c ||  c | c | c |}
\hline
& Vanilla & Pseudo & Thermal & Vanilla & Pseudo & Thermal \\ 
\hline
Dataset &\multicolumn{3}{ c ||}{Planck TT}\\
\cline{1-4}
Planck low-$\ell$ TT & $23.6$ & $21.4$ & $20.2$ \\
Planck low-$\ell$ EE & $395.7$ & $396.4$ & $395.9$ \\
Planck high-$\ell$ TT & $760.6$ & $767.3$ & $774.0$ \\
\cline{1-4}
Total $\chi^2$ & $1180.0$ & $1185.0$ & $1190.1$\\
Total $\Delta \chi^2$ & $0$ & $5$ & $10$ \\
\hline
Dataset &\multicolumn{3}{ c ||}{Planck TTTEEE}&\multicolumn{3}{ c |}{... + R19}\\
\hline
Planck low-$\ell$ TT & $23.5$ & $24.0$ & $20.3$ & $21.5$ & $23.3$ & $20.3$\\
Planck low-$\ell$ EE & $395.8$ & $396.6$ & $396.6$ & $396.9$ & $395.9$ & $397.3$\\
Planck high-$\ell$ TTTEEE & $2346.3$ & $2357.6$ & $2380.5$ & $2355.1$ & $2358.6$ & $2378.5$\\
R19 & $--$ & $--$ & $--$ & $6.4$ & $1.0$ & $0.0$\\
\hline
Total $\chi^2$ & $2765.6$ & $2778.1$ & $2797.3$ & $2779.8$ & $2778.8$ & $2796.0$\\
Total $\Delta \chi^2$ & $0$ & $12.5$ & $31.5$ & $0$ & $-1.0$ & $16.2$\\
\hline
Dataset &\multicolumn{3}{ c ||}{... + lensing + BAO}&\multicolumn{3}{ c |}{... + lensing + BAO + R19}\\
\hline
Planck low-$\ell$ TT & $23.2$ & $26.3$ & $20.7$ & $21.6$ & $24.5$ & $20.5$\\
Planck low-$\ell$ EE & $396.6$ & $395.9$ & $395.9$ & $395.7$ & $395.8$ & $398.6$\\
Planck high-$\ell$ TTTEEE & $2348.5$ & $2358.9$ & $2381.3$ & $2354.1$ & $2358.4$ & $2379.6$\\
Planck lensing & $8.8$ & $9.2$ & $10.8$ & $10.5$ & $11.3$ & $9.8$\\
BAO DR12& $3.8$ & $3.3$ & $3.7$ & $3.5$ & $4.3$ & $3.8$\\
BAO small z& $1.5$ & $1.8$ & $2.4$ & $2.2$ & $2.9$ & $2.5$\\
R19 & $--$ & $--$ & $--$ & $7.5$ & $5.0$ & $0.1$\\
\hline
Total $\chi^2$ & $2782.3$ & $2795.3$ & $2814.7$ & $2795.2$ & $2802.2$ & $2814.8$\\
Total $\Delta \chi^2$ & $0$ & $13.0$ & $32.4$ & $0$ & $7.0$ & $19.6$ \\
\hline
\hline
\end{tabular}
\end{adjustbox}
\caption{Best-fit $\chi^2$ values of each individual dataset for each dataset combination and for each model.
}
\label{tab:chi2}
\end{table}
\begin{figure}[tbp]
\centering 
\begin{tabular}{cc}
\includegraphics[width=.45\linewidth]{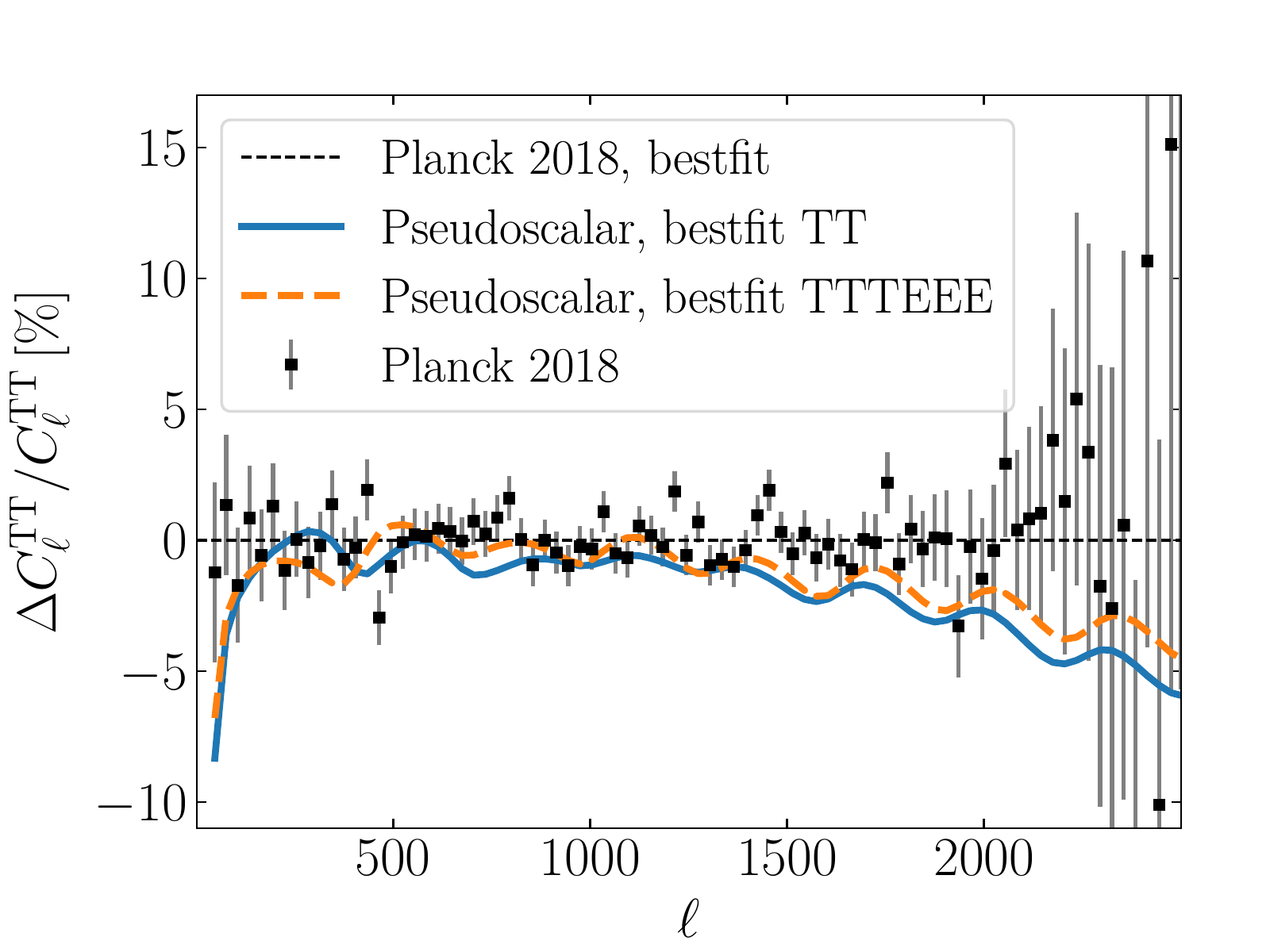}&\includegraphics[width=.45\linewidth]{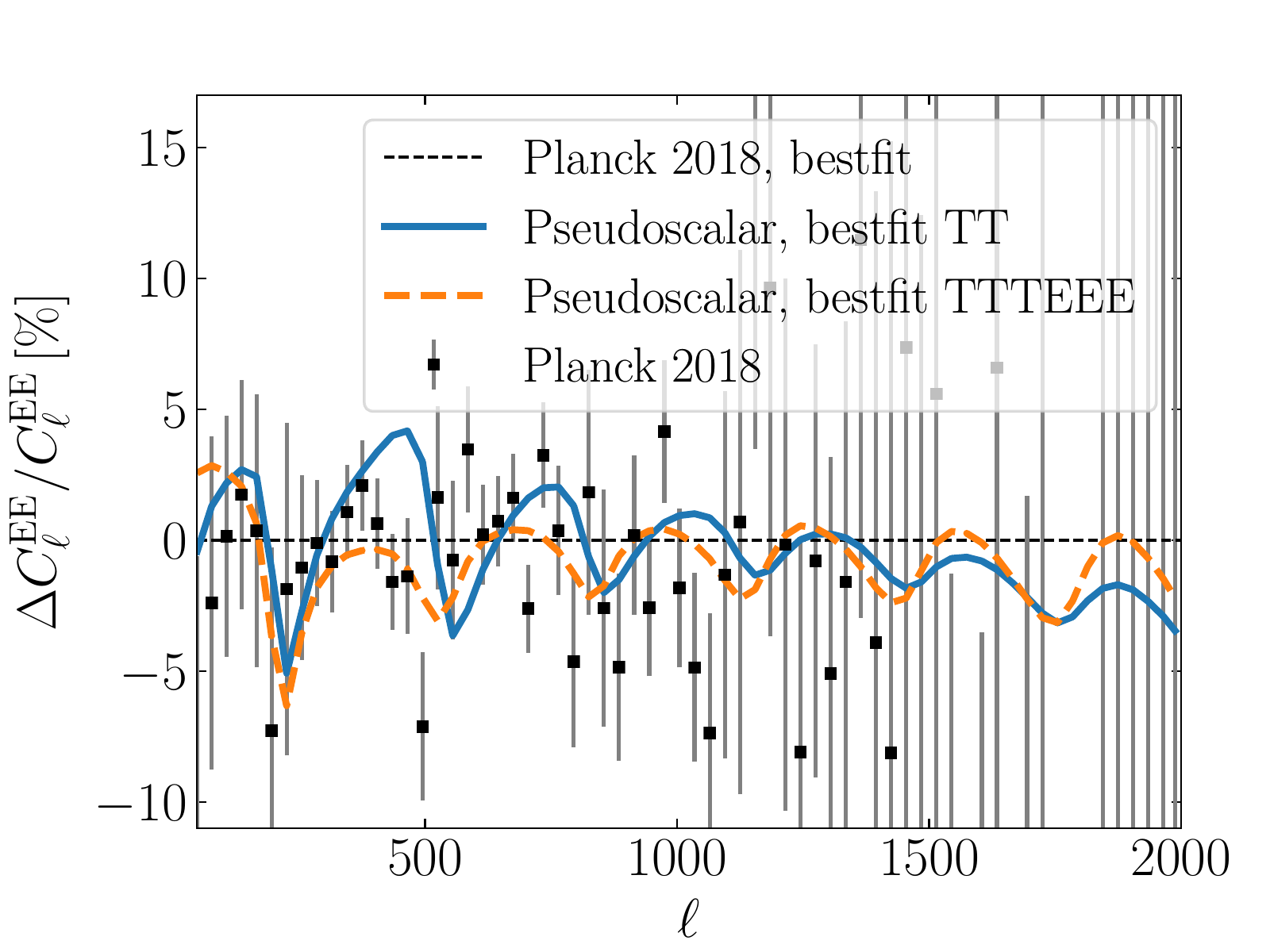}\\
\includegraphics[width=.45\linewidth]{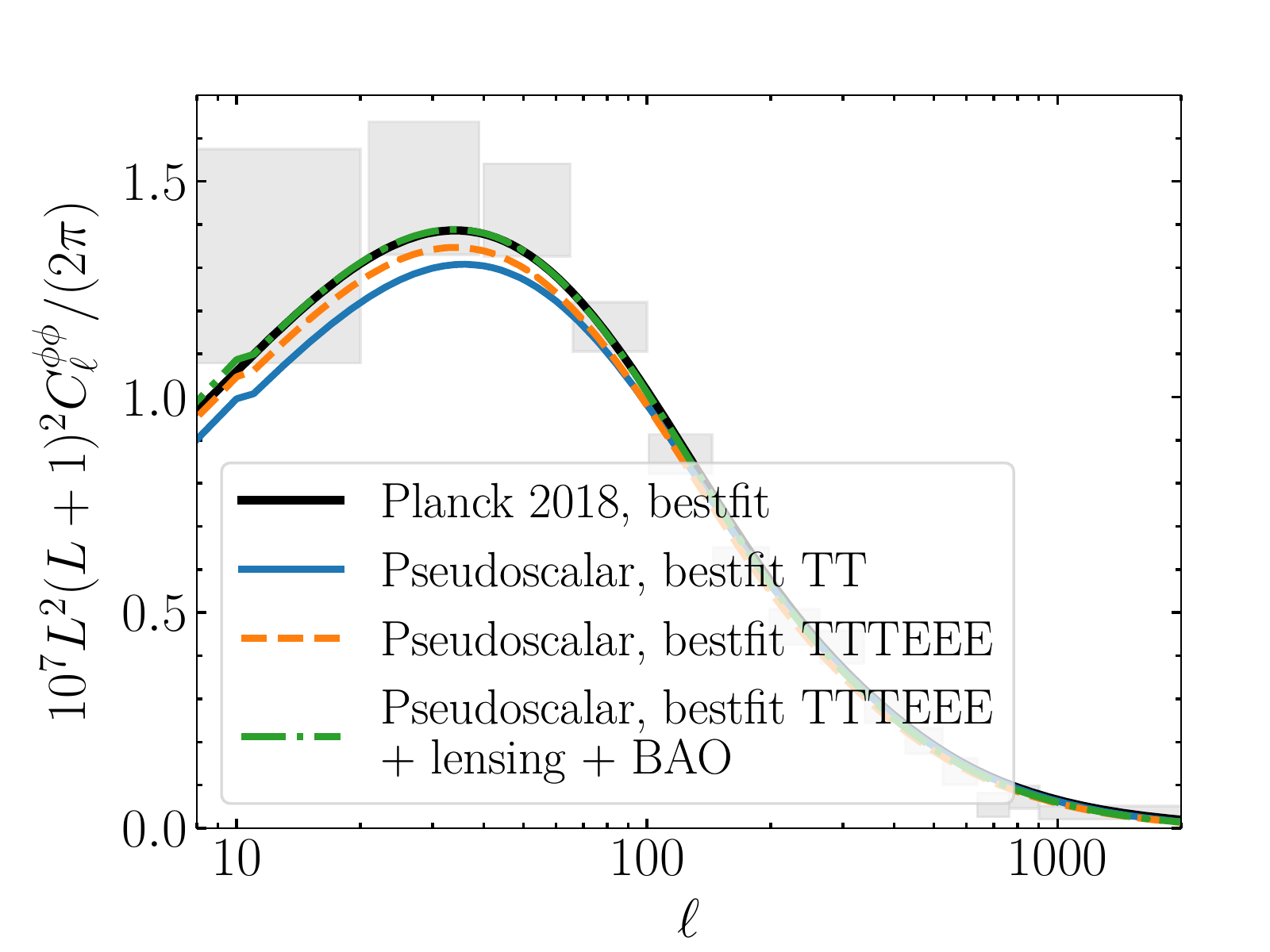}&\includegraphics[width=.45\linewidth]{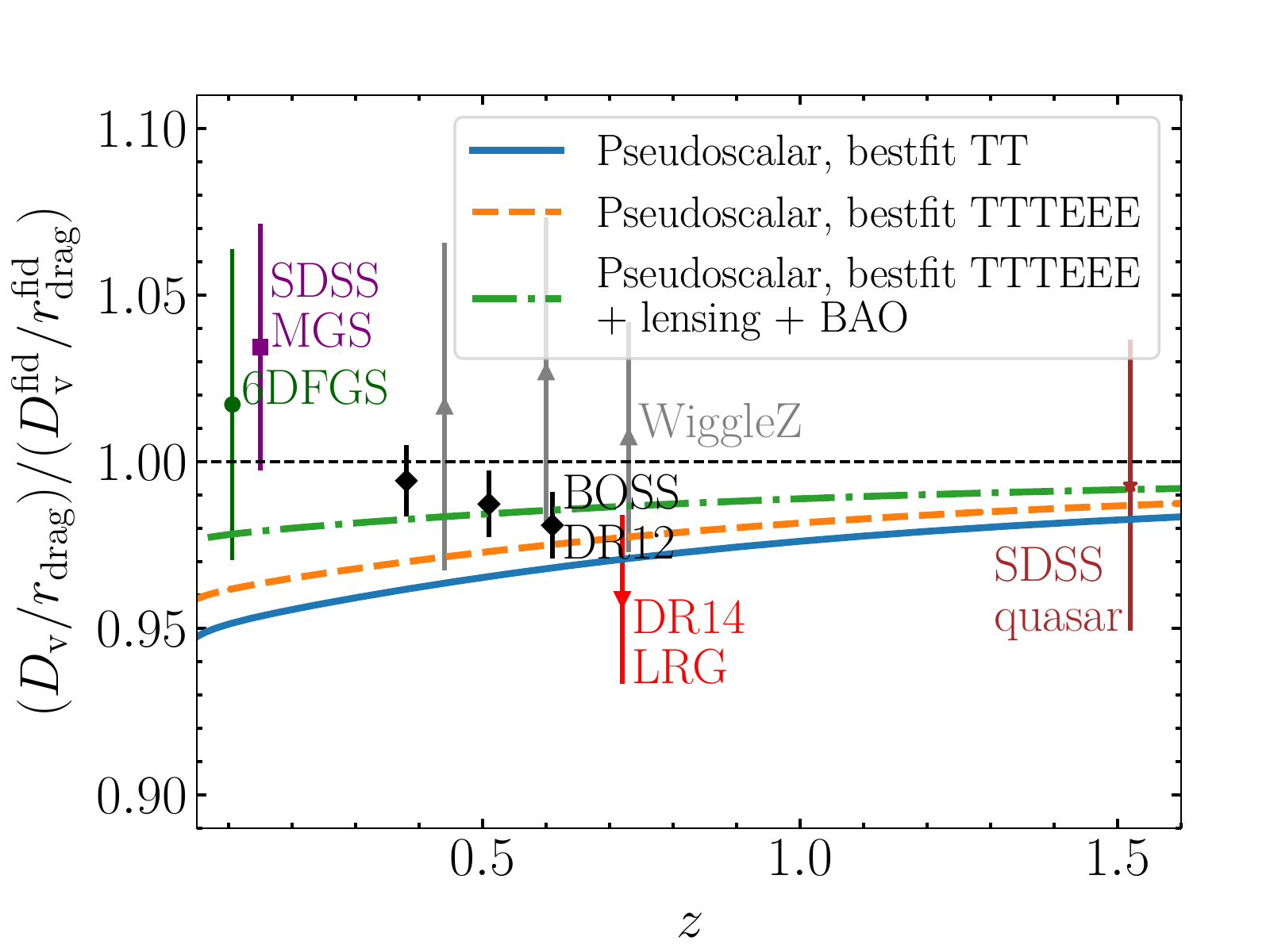}\\
\end{tabular}
\caption{\label{fig:bf} Upper panels: Percentage relative difference in temperature (upper left panel) and polarization (upper right panel) between the Pseudoscalar best-fit of TT (blue solid lines) and of TTTEEE (orange dashed lines) and the Planck 2018 best-fit. The data points with error bars show the Planck 2018 measurements.
Lower left panel: CMB lensing power spectrum of the Planck 2018 best-fit, and Pseudoscalar best-fit of TT (blue solid line), TTTEEE (orange dashed line), and TTTEEE + lensing + BAO (green dot-dashed line). Lower right panel: Ratio between the volume distance ($D_V$) divided by the size of the sound horizon at baryon drag ($r_d$) in the Pseudoscalar best-fit (same colour coding as in the plot of CMB lensing) and in Planck 2018 best-fit. The BAO measurements are: SDSS MGS \cite{Ross:2014qpa}, 6DFGS \cite{Beutler:2011hx}, BOSS DR12 \cite{Alam:2016hwk}, WiggleZ \cite{Kazin:2014qga}, DR14 LRG \cite{Bautista:2017wwp}, SDSS quasar \cite{Blomqvist:2019rah}. Notice that only SDSS MGS, 6DFGS, BOSS DR12 are included in the analysis.
}
\end{figure}

These considerations are confirmed by the $\chi^2$ values in the last three columns of Table \ref{tab:chi2}: the Thermal case perfectly reproduces the local value of $H_0$, thus, the R19 prior has $\chi^2=0$ in such case.
However, the $\chi^2$ values also show that the Thermal case pays the price of a very poor fit of CMB temperature and polarization data, thus, the total $\Delta \chi^2$ with respect to the Vanilla case is still large.
On the contrary, in our Pseudoscalar model the fit of CMB temperature and polarization data, although worse than in the Vanilla scenario (see Figure \ref{fig:bf} upper panels), is still reasonable enough to yield a total $\Delta \chi^2=-1$ in combination with the R19 prior.
Notice that the largest contribution to the Pseudoscalar $\Delta \chi^2$ comes from the high-$\ell$ E-mode polarisation data, which is also responsible for the reduced upper bound on $m_s$ that we already mentioned and that will be further discussed in the next Section.
Concerning the impact of large scale structures, the differences in $\Delta \chi^2$ of Planck TTTEEE + lensing + BAO are pretty much the same as in Planck TTTEEE, for all three models, indicating that they have a similar behaviour at low redshift (see Figure \ref{fig:bf} lower panels). However, in the Pseudoscalar model the fit to BAO data requires a larger value of $\Omega_m$, which in turns leads to a smaller value of both $\Delta N_{\rm eff}$ and $H_0$. As a consequence, the consistency with R19 is reduced in the combined fit Planck TTTEEE + lensing + BAO + R19.

\subsection{CMB phenomenology in the pseudoscalar model}
\label{sec:phenomenology}

\begin{figure}[tbp]
\includegraphics[width=\linewidth]{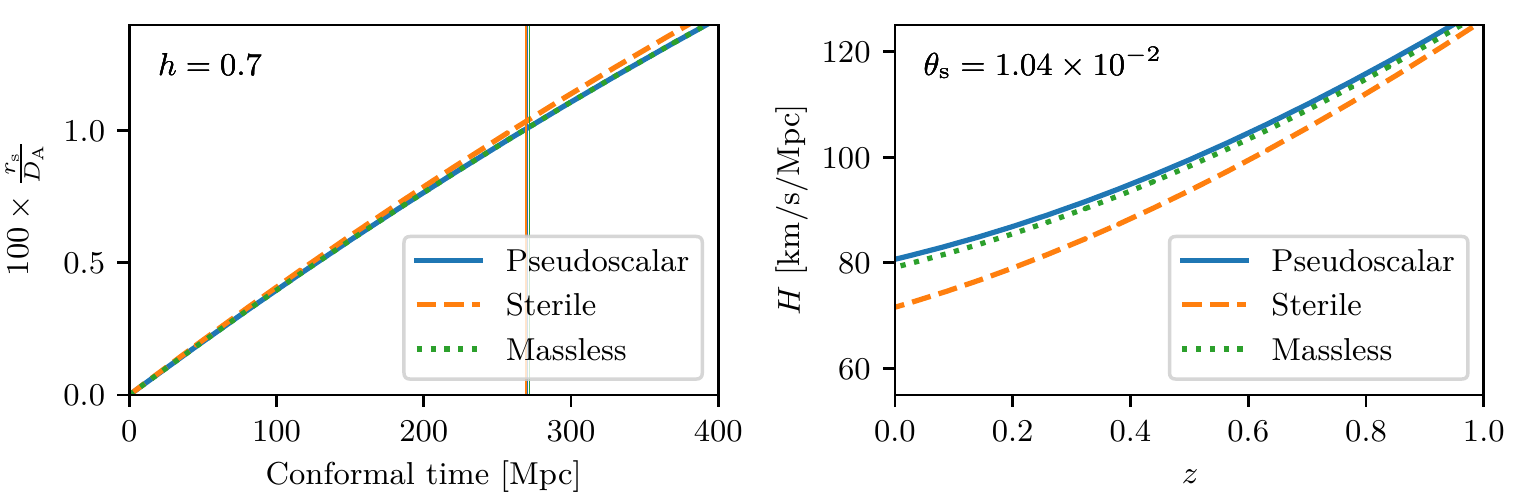}
\caption{\label{fig:rsDAH}
\emph{Left panel:} The effective angular sound horizon as a function of conformal time in the early Universe with $h=0.7$ for the three cases ``Pseudoscalar'', ``Sterile'' and ``Massless'' described in the text. The vertical lines mark the conformal time of recombination.
\emph{Right panel:} Hubble rate for the same models in the late Universe with fixed angular scale of the sound-horizon at recombination $100\times\theta_s=1.04$.
}
\end{figure}

In order to better understand why the addition of high-$l$ polarisation data so severely constrains the mass of the sterile neutrino we have investigated the effect of the pseudoscalar model on CMB anisotropies in more detail.
In the left panel of figure~\ref{fig:rsDAH} we show the angular scale of the sound horizon, $\theta_s$, for three different cases: 

\begin{description}
\item[Pseudoscalar:] The full Pseudoscalar model with $N_\mathrm{pseudo}=1$ and $m_s=1$~eV.
\item[Massless:] The Thermal model with one massless sterile neutrino, corresponding to $N_\nu=4.046$ massless, non-interacting neutrinos.
\item[Sterile:] The Thermal model with one massive sterile neutrino, with the same mass as in the Pseudoscalar model ($m_s=1$~eV), but no interactions.
\end{description}

\begin{figure}[tbp]
\includegraphics[width=\linewidth]{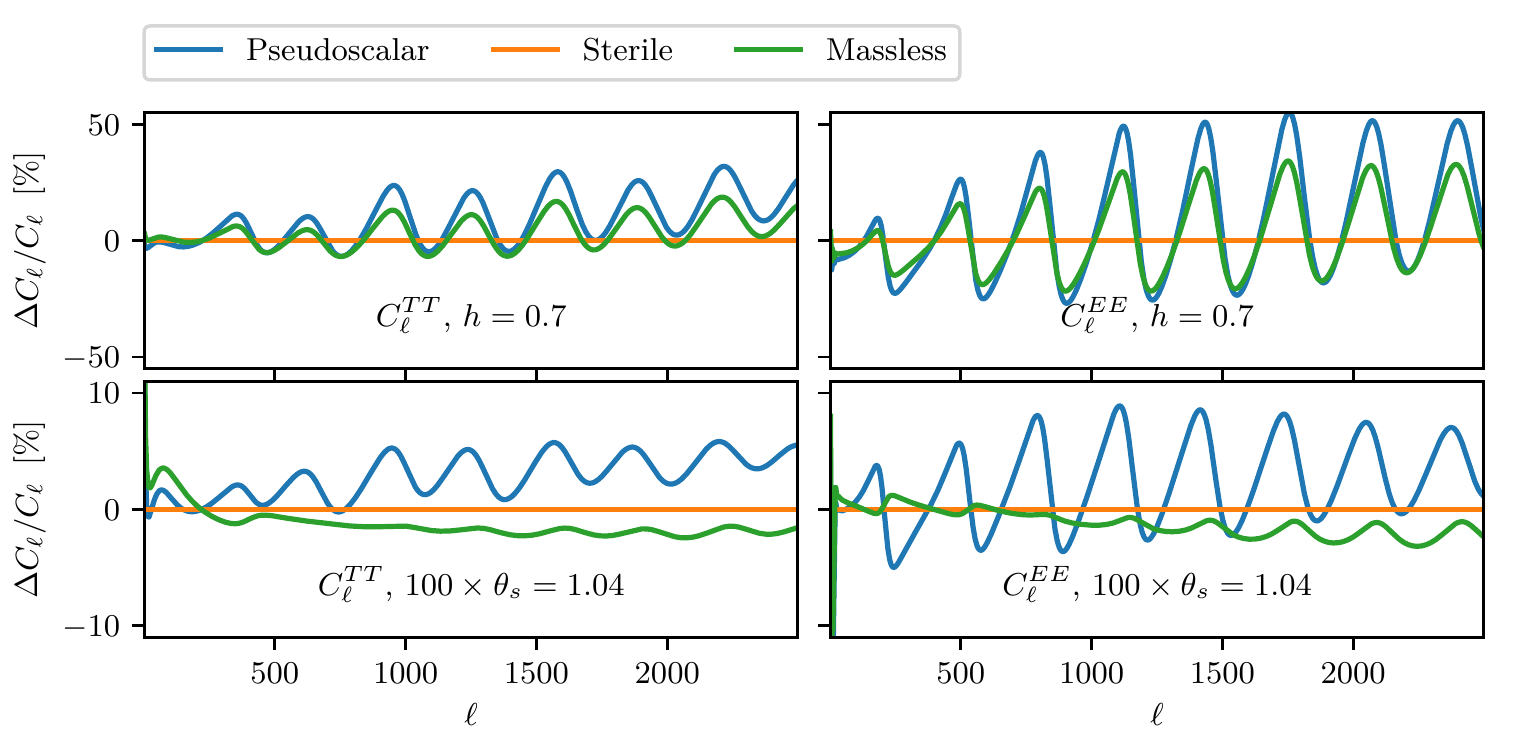}
\caption{
\label{fig:TTEE}
Relative differences in $C_\ell^{TT}$ (left panels) and $C_\ell^{EE}$ (right panels). The upper panels have $h=0.7$, and the ``Sterile'' case is clearly more different due to its different late-time behaviour. The lower panels show the case where the angular sound-horizon at recombination has been fixed, and this reduces the scatter by a factor 5.
Now the ``Massless'' and the ``Sterile'' case are much more similar while the ``Pseudoscalar'' case stands out. 
}
\end{figure}

Let us first consider the case where the value of the Hubble parameter is $h=0.7$.
As can be seen from the left panel of Figure~\ref{fig:rsDAH}, the Sterile case has a different angular scale of the sound-horizon due to the additional late-time energy density.
In the upper panels of Figure~\ref{fig:TTEE} we show the corresponding relative difference in the $C_\ell^{TT}$ and $C_\ell^{EE}$ spectra for fixed $h=0.7$.
The differences of both the Massless and the Pseudoscalar are oscillating, indicating that the peak-structure of the models does not align.
In order to better compare the impact on observables we may also fix the angular scale of the sound-horizon at recombination, $\theta_s$, to be identical in the three cases (which then leads to different values of $h$). The Hubble rate as a function of redshift for the three cases is shown in the right panel of Figure~\ref{fig:rsDAH}, and the resulting relative differences in CMB spectra are shown in the lower panels of Figure~\ref{fig:TTEE}. Now, while the residual deviation of the Massless is within the observational error, the Pseudoscalar case stands out in two different ways:
\begin{enumerate}
\item There is a residual horizontal peak-shift in both $C_\ell^{TT}$ and $C_\ell^{EE}$ even after $\theta_s$ has been fixed.
\item The $C_\ell^{TT}$ spectrum shows a coherent increase in power at large $\ell$ compared to the other two cases. 
\end{enumerate}
The first is a non-trivial phase-shift of the acoustic oscillations in the photon-baryon plasma due to the fact that in the Pseudoscalar model the sterile sector behaves as fluid rather than a free-streaming component (see e.g.~\cite{Pan:2016zla,Bashinsky:2003tk,Baumann:2015rya} for a discussion of the phase-shift effect). The second effect is a reduction in Silk damping at fixed $\theta_s$ due to the different expansion histories which is also responsible for the well-known $n_s$--$N_\text{eff}$ correlation~\cite{Hou:2011ec}.

\begin{figure}[tbp]
\centering 
\includegraphics[width=.75\linewidth]{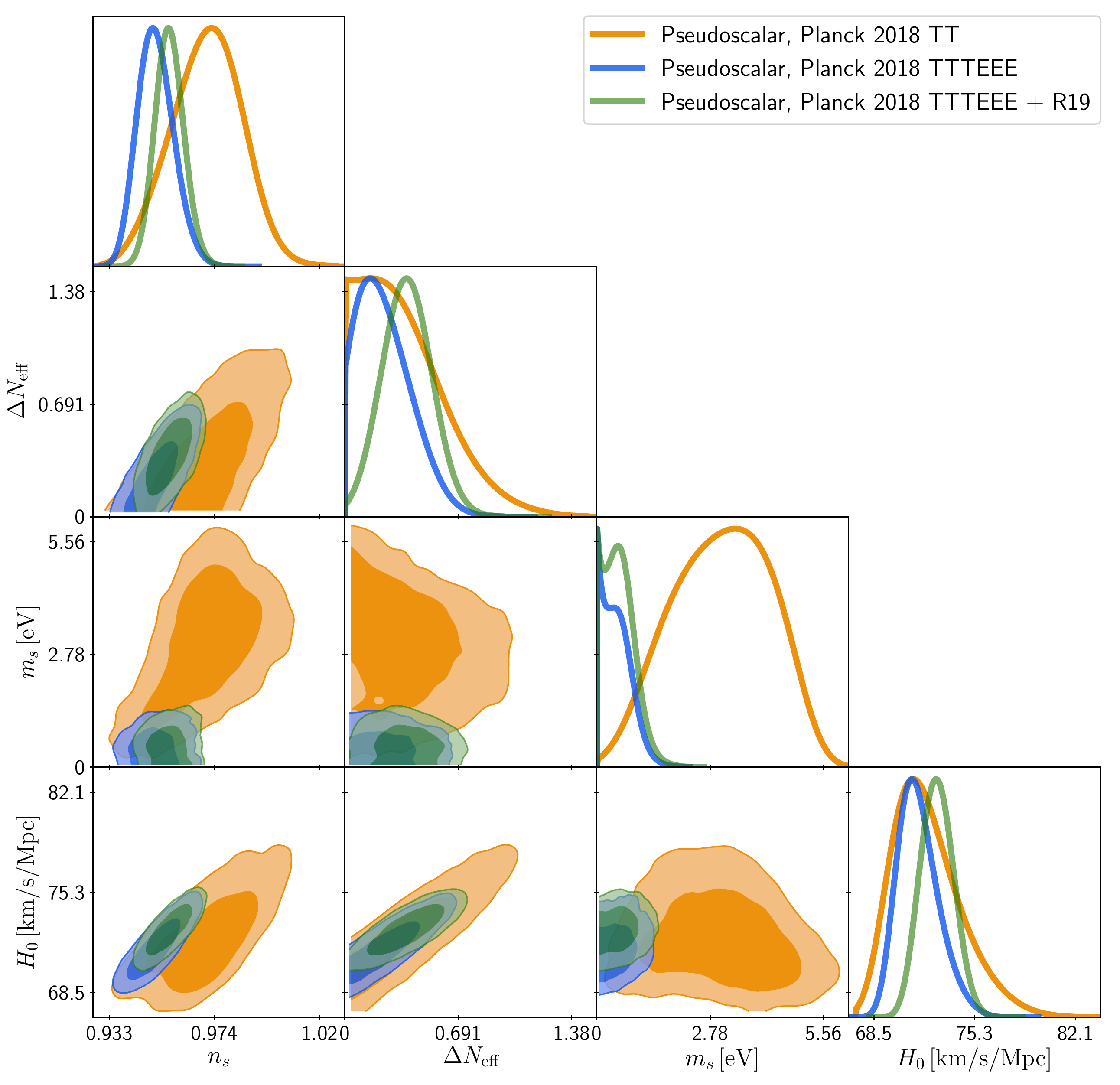}
\caption{\label{fig:ns} Marginalized 2D $1\,\sigma$ (dim) and $2\,\sigma$ (light) contours and 1D posterior for a subset of cosmological parameters $(n_s,\,\Delta N_{\rm eff},\,m_s,\,H_0)$.
}
\end{figure}

When only $C_\ell^{TT}$-data is included, the second effect may be compensated by a change in $n_s$. We can see this in Figure~\ref{fig:ns}, where it is clear that there is a relatively strong correlation between $n_s$, $m_s$, and $\Delta N_{\rm eff}$. Once polarisation is included, $n_s$ is much more tightly constrained and this in turn severely restricts both $m_s$ and $\Delta N_{\rm eff}$.

It is now clear why the sterile neutrino mass becomes tightly constrained when polarisation data is added, even in the pseudoscalar model where there are no direct late-time effects of the mass.

\section{Short-baseline neutrino oscillations}
\label{sec:SBL}

It is interesting to investigate the implications of the cosmological constraints on the
sterile neutrino mass for short-baseline (SBL) neutrino oscillations.
Several SBL neutrino oscillation experiments are operating or in preparation
(see the recent reviews in Refs.~\cite{Giunti:2019aiy,Diaz:2019fwt,Boser:2019rta}),
motivated by the following three well-known anomalies:
1) the LSND observation of short-baseline $\bar\nu_{\mu}\to\bar\nu_{e}$ transitions \cite{Aguilar:2001ty}
(the LSND anomaly),
somewhat supported by the results of the MiniBooNE
experiment~\cite{Aguilar-Arevalo:2018gpe};
2) the indication of short-baseline $\nu_e$ disappearance
\cite{Laveder:2007zz,Giunti:2006bj}
in the
GALLEX
\cite{Kaether:2010ag}
and
SAGE
\cite{Abdurashitov:2005tb}
gallium source experiments
(the gallium neutrino anomaly);
3) the short-baseline $\bar\nu_e$ disappearance of reactor $\bar\nu_e$
\cite{Mention:2011rk}
with respect to the theoretical prediction of the reactor $\bar\nu_e$ fluxes
\cite{Mueller:2011nm,Huber:2011wv}
(the reactor antineutrino anomaly).
Recent global fits of SBL oscillation data in terms of 3+1 active-sterile mixing
found a strong appearance-disappearance tension~\cite{Gariazzo:2018mwd,Dentler:2018sju},
mainly due to the stringent MINOS/MINOS+ limits on $\nu_{\mu}$ disappearance
combined with the reactor limits on $\nu_{e}$ disappearance.
However,
the oscillation explanation of the SBL anomalies still cannot be dismissed,
taking into account that it is the only general one that avoids
a multitude of different exotic ad-hoc solutions of each experimental anomaly.
It is possible that the appearance-disappearance tension
is due to a misleading interpretation of the data of one or more experiments,
for which some systematic effect has been overlooked.
Moreover,
the experiments searching for SBL neutrino oscillations need an
indication of the most likely region of the oscillation parameter space
for tuning their sensitivities.
Therefore, it is useful to perform a global fit of the SBL data
in the 3+1 active-sterile mixing in spite of the appearance-disappearance tension.
This strategy is also in agreement with the Bayesian philosophy
that considers the experimental observations as means to improve our knowledge on a model.
A model can be rejected only if there is a better alternative.
Since at present we do not have a general alternative model
that can explain the SBL anomalies,
it is appropriate to increase our Bayesian knowledge of
3+1 active-sterile mixing by performing a Bayesian global analysis of the
SBL neutrino oscillation data that can be compared and combined with
Bayesian analyses of cosmological data.

In our analysis we considered the following sets of SBL oscillation data:
\begin{description}

\item[$\nu_{e}$ disappearance:]
The ratio of the spectra measured at different distance from the source
in the
Bugey-3~\cite{Declais:1994su},
NEOS~\cite{Ko:2016owz},
DANSS~\cite{Danilov:2019aef}, and
PROSPECT~\cite{Ashenfelter:2018iov}
reactor neutrino experiments.
The ratio of the
KARMEN~\cite{Armbruster:1998uk}
and
LSND~\cite{Auerbach:2001hz}
$\nu_{e} + {}^{12}\text{C} \to {}^{12}\text{N}_{\text{g.s.}} + e^{-}$
scattering data at 18~m and 30~m from the source
\cite{Conrad:2011ce,Giunti:2011cp}.
The solar neutrino bound~\cite{Gariazzo:2017fdh}.

\item[$\nu_{\mu}$ disappearance:]
The constraints from the analyses of the data of the
CDHSW~\cite{Dydak:1983zq},
CCFR~\cite{Stockdale:1984cg},
SciBooNE-MiniBooNE~\cite{Mahn:2011ea,Cheng:2012yy},
IceCube~\cite{TheIceCube:2016oqi}\footnote{
We cannot take into account of the new interesting results of the IceCube
experiment~\cite{2005.12942,2005.12943}
that appeared in arXiv during the completion of this work.
},
MINOS/MINOS+~\cite{Adamson:2017uda}, and
atmospheric~\cite{Maltoni:2007zf}
neutrino experiments.

\item[$\nu_{\mu}\to\nu_{e}$ appearance:]
The constraints of the
BNL-E776~\cite{Borodovsky:1992pn},
KARMEN~\cite{Armbruster:2002mp},
NOMAD~\cite{Astier:2003gs},
ICARUS~\cite{Antonello:2013gut}
and
OPERA~\cite{Agafonova:2013xsk}
experiments.
The LSND data~\cite{Aguilar:2001ty} with the
$\chi^2$ map calculated by the LSND collaboration
taking into account the decay-at-rest and the decay-in-flight data.
The MiniBooNE data~\cite{Aguilar-Arevalo:2018gpe}
according to the official data release in \url{https://www-boone.fnal.gov/for_physicists/data_release/nue2018/}.

\end{description}

\begin{figure}
\centering 
\includegraphics[width=.45\linewidth]{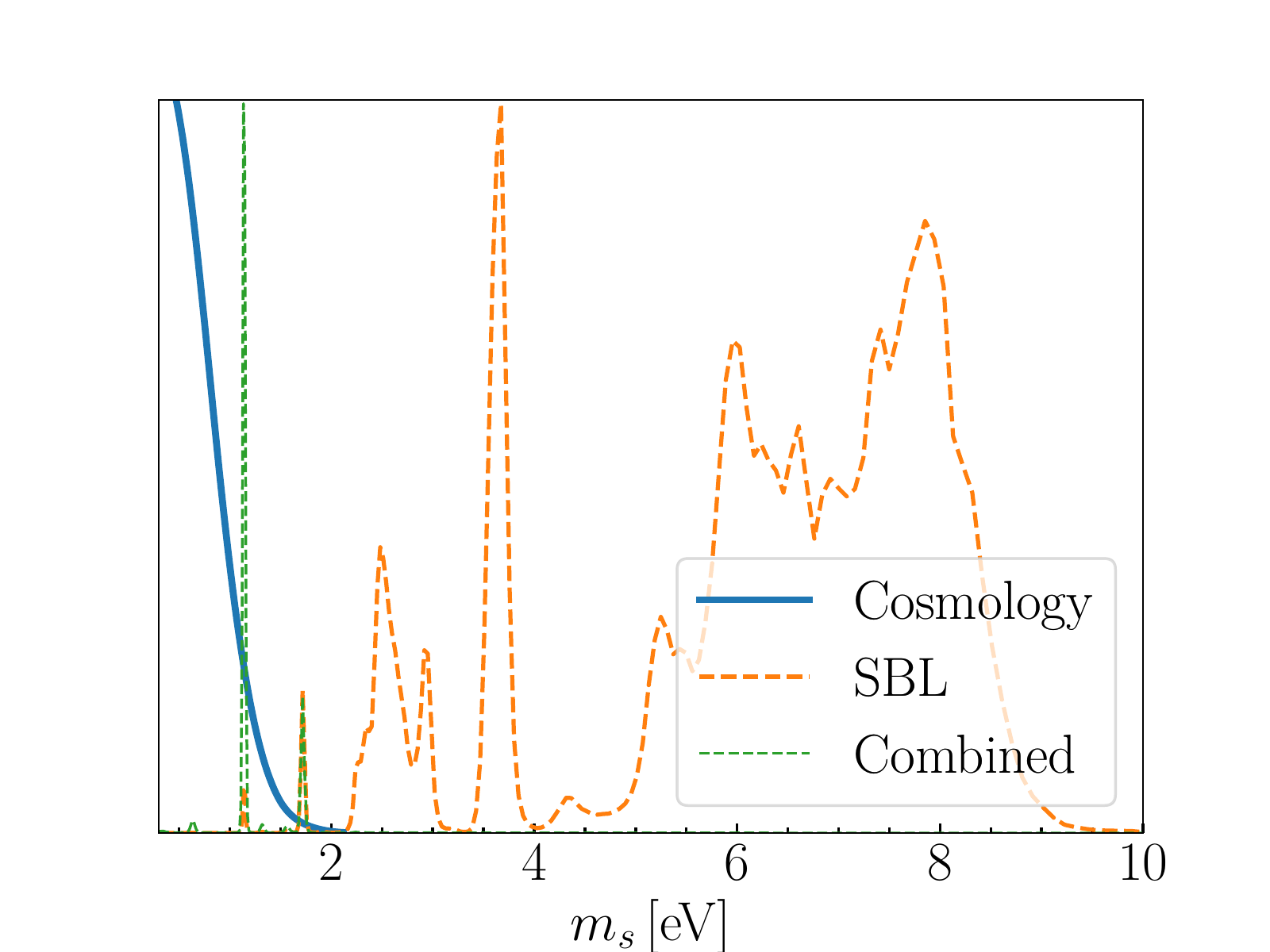}
\caption{
\label{fig:ms}
Marginalized 1D posteriors of $m_s$, for SBL, cosmology (corresponding to the Pseudoscalar model fit of Planck 2018 TTTEEE), and for the combined analysis.}
\end{figure}

We calculated the posterior probability for $m_{s}$ shown in Fig.~\ref{fig:ms}.
One can see that there are significant peaks of the SBL probability (orange)
for $m_{s} \simeq 1.14 \, \text{eV}$
and $m_{s} \simeq 1.72 \, \text{eV}$,
but larger values of $m_{s}$ are favored by the SBL data.
Since the posterior obtained considering the Pseudoscalar model and Planck TTTEEE data (blue)
has some overlapping with the peaks at small $m_{s}$,
it is useful to combine the results of the cosmological and SBL analyses
in order to restrict the allowed range of $m_{s}$.
The posterior probability of the combined fit (green) is maximal at $m_{s} \simeq 1.14 \, \text{eV}$
and has a significant local maximum at $m_{s} \simeq 1.72 \, \text{eV}$,
in correspondence of the local peaks of the SBL probability mentioned above.
There are also minor peaks at
$m_{s} \simeq 0.64 \, \text{eV}$,
$m_{s} \simeq 1.32 \, \text{eV}$, and
$m_{s} \simeq 1.52 \, \text{eV}$.
On the other hand, the combination with the cosmological constraints
disfavors all the peaks at higher masses that are allowed by oscillation data alone.
We verified that the situation is approximately the same if instead of the posterior obtained
with Planck TTTEEE data alone we also include a prior on $H_0$.
We conclude that the Pseudoscalar model still allows to find a reasonable fit
to SBL and cosmological data together,
and the best-fit of the sterile neutrino mass remains around $1$ eV.

\section{Conclusions}
\label{sec:conclusions}

We presented an updated analysis of how light sterile neutrino parameters can be constrained using cosmological data.
The simplest case we studied is the one we refer to as ``Thermal'', in which the additional sterile state is assumed to be completely thermalised prior to active neutrino decoupling so that it has the same temperature as the active species.
This would be e.g.\ the case for the masses and mixings required for a sterile neutrino explanation of the observed SBL anomalies.
In this case we find that the sterile mass is constrained to be below 0.2-0.9~eV, depending on the combination of data used.
Consistent with other recent analyses we found that, although the presence of a fully thermalised extra neutrino pushes the value of $H_0$ inferred from CMB up, making it consistent with local measurements, it comes at the cost of a much poorer fit of CMB data.
Without the inclusion of the R19~\cite{Riess:2019cxk} Hubble data, the $\Delta \chi^2$ is larger than 30 compared to the standard $\Lambda$CDM. Even when the R19 prior is included the thermal case is disfavoured by $\Delta \chi^2 \sim 16$.

The second case we considered is one where the contribution of steriles to $N_{\rm eff}$ is allowed to vary (we refer to this as the ``Vanilla'' case).
Not surprisingly we recover the well-known warm dark matter limit in which the physical mass is poorly constrained because a small value of $\Delta N_{\rm eff}$ allows for a very large sterile neutrino mass.
Furthermore, in this case the preferred value of $H_0$ reverts to the $\Lambda$CDM value because $\Delta N_{\rm eff}$ is allowed to be small.

Finally, we considered the model studied in \cite{Archidiacono:2014nda,Archidiacono:2015oma,Archidiacono:2016kkh}, in which the steriles are charged under a new pseudoscalar interaction.
In this case the sterile neutrinos and pseudoscalars form a strongly self-interacting fluid long before photon decoupling, but after neutrino decoupling.
The model provides a specific prediction for the ratio of energy density in the sterile+pseudoscalar sector to the energy density in standard model neutrinos.
However, the total energy density in the two components can be modified by e.g.\ early production of steriles or pseudoscalars, and we therefore treated $\Delta N_{\rm eff}$ as a free parameter in this case.
The Pseudoscalar model is qualitatively different from models with non-interacting steriles, both because of the lack of free-streaming in the sterile+pseudoscalar component, {\it and} because of the annihilation of steriles to pseudoscalars when the temperature drops below the mass of the sterile.
In the Pseudoscalar model we found that when only CMB temperature measurements are used the mass bound essentially disappears.
This is completely consistent with an earlier analysis using a previous Planck data release \cite{Archidiacono:2016kkh}.
However, once polarisation data at high $\ell$ is added, a large part of the sterile neutrino parameter space is excluded, because the degeneracy between $m_s, \Delta N_{\rm eff},$ and $n_s$ is broken.
This leads to a cosmological upper bound in the 1~eV range. 
Consistent with previous analyses, we still find that the Pseudoscalar model is in good agreement with the local $H_0$ measurement, since it predicts a value of $H_0$ around 72 ${\rm km} \, {\rm s}^{-1} \, {\rm Mpc}^{-1}$.
As a consequence, the combined CMB+R19 fit is slightly better than the one obtained for the simpler $\Lambda$CDM model.
This mild preference in favour of the Pseudoscalar model disappears once CMB lensing and Baryonic Acoustic Oscillations are included: although the model is consistent with BAO and CMB lensing, with a $\chi^2$ similar to the one of $\Lambda$CDM, the value of the Hubble constant is shifted towards smaller values, due to the $\Omega_m\,-\,H_0$ anti-correlation, hence the consistency with R19 is reduced. Nevertheless, the $H_0$ tension is alleviated in the Pseudoscalar model with respect to $\Lambda$CDM.

Finally, we also performed an analysis of available short baseline data in combination with cosmology for the pseudoscalar model.
Contrary to our previous analyses, the addition of high-$\ell$ polarisation data now excludes a large part of the mass range suggested by SBL data.
This is shown in Fig.~\ref{fig:ms},
where one can see that the combined posterior now singles out narrow intervals around 1~eV as the only viable mass ranges for the sterile neutrino.

\acknowledgments
C.G.\ would like to thank W.C. Louis for kindly providing the LSND $\chi^2$ map.
S.G.\ acknowledges partial financial support by the ``Juan de la Cierva-Incorporaci\'on'' program (IJC2018-036458-I) of the Spanish MINECO, and from the Spanish grants FPA2017-85216-P (AEI/FEDER, UE), PROMETEO/2018/165 (Generalitat Valenciana) and the Red Consolider MultiDark FPA2017-90566-REDC, until September 2020. Starting from October 2020, he has received funding from the European Union's Horizon 2020 research and innovation programme under the Marie Skłodowska-Curie grant agreement No 754496 (project FELLINI).
The work of C.G.\ was partially supported by the research grant ``The Dark Universe: A Synergic Multimessenger Approach'' number 2017X7X85K under the program PRIN 2017 funded by the Ministero dell'Istruzione, Universit\`a e della Ricerca (MIUR).
T.T.\ was supported by a research grant (29337) from VILLUM FONDEN.

\bibliography{references}{}

\providecommand{\href}[2]{#2}\begingroup\raggedright\begin{thebibliography}{100}

\bibitem{Aghanim:2018eyx}
{\scshape Planck} collaboration, N.~Aghanim et~al., \emph{{Planck 2018 results.
  VI. Cosmological parameters}},  \href{https://arxiv.org/abs/1807.06209}{{\tt
  1807.06209}}.

\bibitem{Riess:2020sih}
A.~G. Riess, \emph{{The Expansion of the Universe is Faster than Expected}},
  \href{http://dx.doi.org/10.1038/s42254-019-0137-0}{\emph{Nature Rev. Phys.}
  {\bf 2} (2019) 10--12}, [\href{https://arxiv.org/abs/2001.03624}{{\tt
  2001.03624}}].

\bibitem{Verde:2019ivm}
L.~Verde, T.~Treu and A.~G. Riess, \emph{{Tensions between the Early and the
  Late Universe}},  in \emph{{Nature Astronomy 2019}}, 2019.
\newblock \href{https://arxiv.org/abs/1907.10625}{{\tt 1907.10625}}.
\newblock \href{http://dx.doi.org/10.1038/s41550-019-0902-0}{DOI}.

\bibitem{Riess:2019cxk}
A.~G. Riess, S.~Casertano, W.~Yuan, L.~M. Macri and D.~Scolnic, \emph{{Large
  Magellanic Cloud Cepheid Standards Provide a 1\% Foundation for the
  Determination of the Hubble Constant and Stronger Evidence for Physics beyond
  $\Lambda$CDM}},
  \href{http://dx.doi.org/10.3847/1538-4357/ab1422}{\emph{Astrophys. J.} {\bf
  876} (2019) 85}, [\href{https://arxiv.org/abs/1903.07603}{{\tt 1903.07603}}].

\bibitem{Huang:2019yhh}
C.~D. Huang, A.~G. Riess, W.~Yuan, L.~M. Macri, N.~L. Zakamska, S.~Casertano
  et~al., \emph{{Hubble Space Telescope Observations of Mira Variables in the
  Type Ia Supernova Host NGC 1559: An Alternative Candle to Measure the Hubble
  Constant}},  \href{https://arxiv.org/abs/1908.10883}{{\tt 1908.10883}}.

\bibitem{Yuan:2019npk}
W.~Yuan, A.~G. Riess, L.~M. Macri, S.~Casertano and D.~Scolnic,
  \emph{{Consistent Calibration of the Tip of the Red Giant Branch in the Large
  Magellanic Cloud on the Hubble Space Telescope Photometric System and a
  Re-determination of the Hubble Constant}},
  \href{http://dx.doi.org/10.3847/1538-4357/ab4bc9}{\emph{Astrophys. J.} {\bf
  886} (2019) 61}, [\href{https://arxiv.org/abs/1908.00993}{{\tt 1908.00993}}].

\bibitem{Abbott:2017xzu}
{\scshape LIGO Scientific, Virgo, 1M2H, Dark Energy Camera GW-E, DES, DLT40,
  Las Cumbres Observatory, VINROUGE, MASTER} collaboration, B.~P. Abbott
  et~al., \emph{{A gravitational-wave standard siren measurement of the Hubble
  constant}}, \href{http://dx.doi.org/10.1038/nature24471}{\emph{Nature} {\bf
  551} (2017) 85--88}, [\href{https://arxiv.org/abs/1710.05835}{{\tt
  1710.05835}}].

\bibitem{Jimenez:2001gg}
R.~Jimenez and A.~Loeb, \emph{{Constraining cosmological parameters based on
  relative galaxy ages}},
  \href{http://dx.doi.org/10.1086/340549}{\emph{Astrophys. J.} {\bf 573} (2002)
  37--42}, [\href{https://arxiv.org/abs/astro-ph/0106145}{{\tt
  astro-ph/0106145}}].

\bibitem{Moresco:2016mzx}
M.~Moresco, L.~Pozzetti, A.~Cimatti, R.~Jimenez, C.~Maraston, L.~Verde et~al.,
  \emph{{A 6\% measurement of the Hubble parameter at $z\sim0.45$: direct
  evidence of the epoch of cosmic re-acceleration}},
  \href{http://dx.doi.org/10.1088/1475-7516/2016/05/014}{\emph{JCAP} {\bf 1605}
  (2016) 014}, [\href{https://arxiv.org/abs/1601.01701}{{\tt 1601.01701}}].

\bibitem{Wong:2019kwg}
K.~C. Wong et~al., \emph{{H0LiCOW XIII. A 2.4\% measurement of $H_{0}$ from
  lensed quasars: $5.3\sigma$ tension between early and late-Universe probes}},
   \href{https://arxiv.org/abs/1907.04869}{{\tt 1907.04869}}.

\bibitem{Pesce:2020xfe}
D.~W. Pesce et~al., \emph{{The Megamaser Cosmology Project. XIII. Combined
  Hubble constant constraints}},
  \href{http://dx.doi.org/10.3847/2041-8213/ab75f0}{\emph{Astrophys. J.} {\bf
  891} (2020) L1}, [\href{https://arxiv.org/abs/2001.09213}{{\tt 2001.09213}}].

\bibitem{Schombert:2020pxm}
J.~Schombert, S.~McGaugh and F.~Lelli, \emph{{Using The Baryonic Tully-Fisher
  Relation to Measure $H_o$}},  \href{https://arxiv.org/abs/2006.08615}{{\tt
  2006.08615}}.

\bibitem{Bernal:2018cxc}
J.~Luis~Bernal and J.~A. Peacock, \emph{{Conservative cosmology: combining data
  with allowance for unknown systematics}},
  \href{http://dx.doi.org/10.1088/1475-7516/2018/07/002}{\emph{JCAP} {\bf 1807}
  (2018) 002}, [\href{https://arxiv.org/abs/1803.04470}{{\tt 1803.04470}}].

\bibitem{Knox:2019rjx}
L.~Knox and M.~Millea, \emph{{Hubble constant hunter's guide}},
  \href{http://dx.doi.org/10.1103/PhysRevD.101.043533}{\emph{Phys. Rev.} {\bf
  D101} (2020) 043533}, [\href{https://arxiv.org/abs/1908.03663}{{\tt
  1908.03663}}].

\bibitem{Smith:2019ihp}
T.~L. Smith, V.~Poulin and M.~A. Amin, \emph{{Oscillating scalar fields and the
  Hubble tension: a resolution with novel signatures}},
  \href{https://arxiv.org/abs/1908.06995}{{\tt 1908.06995}}.

\bibitem{Lemos:2018smw}
P.~Lemos, E.~Lee, G.~Efstathiou and S.~Gratton, \emph{{Model independent $H(z)$
  reconstruction using the cosmic inverse distance ladder}},
  \href{http://dx.doi.org/10.1093/mnras/sty3082}{\emph{Mon. Not. Roy. Astron.
  Soc.} {\bf 483} (2019) 4803--4810},
  [\href{https://arxiv.org/abs/1806.06781}{{\tt 1806.06781}}].

\bibitem{Benevento:2020fev}
G.~Benevento, W.~Hu and M.~Raveri, \emph{{Can Late Dark Energy Transitions
  Raise the Hubble constant?}},  \href{https://arxiv.org/abs/2002.11707}{{\tt
  2002.11707}}.

\bibitem{Dhawan:2020xmp}
S.~Dhawan, D.~Brout, D.~Scolnic, A.~Goobar, A.~Riess and V.~Miranda,
  \emph{{Cosmological model insensitivity of local $H_0$ from the Cepheid
  distance ladder}},  \href{https://arxiv.org/abs/2001.09260}{{\tt
  2001.09260}}.

\bibitem{Hamann:2010bk}
J.~Hamann, S.~Hannestad, G.~G. Raffelt, I.~Tamborra and Y.~Y.~Y. Wong,
  \emph{{Cosmology seeking friendship with sterile neutrinos}},
  \href{http://dx.doi.org/10.1103/PhysRevLett.105.181301}{\emph{Phys. Rev.
  Lett.} {\bf 105} (2010) 181301}, [\href{https://arxiv.org/abs/1006.5276}{{\tt
  1006.5276}}].

\bibitem{Motohashi:2012wc}
H.~Motohashi, A.~A. Starobinsky and J.~Yokoyama, \emph{{Cosmology Based on f(R)
  Gravity Admits 1 eV Sterile Neutrinos}},
  \href{http://dx.doi.org/10.1103/PhysRevLett.110.121302}{\emph{Phys. Rev.
  Lett.} {\bf 110} (2013) 121302}, [\href{https://arxiv.org/abs/1203.6828}{{\tt
  1203.6828}}].

\bibitem{Giusarma:2011zq}
E.~Giusarma, M.~Archidiacono, R.~de~Putter, A.~Melchiorri and O.~Mena,
  \emph{{Sterile neutrino models and nonminimal cosmologies}},
  \href{http://dx.doi.org/10.1103/PhysRevD.85.083522}{\emph{Phys. Rev.} {\bf
  D85} (2012) 083522}, [\href{https://arxiv.org/abs/1112.4661}{{\tt
  1112.4661}}].

\bibitem{Hamann:2011ge}
J.~Hamann, S.~Hannestad, G.~G. Raffelt and Y.~Y.~Y. Wong, \emph{{Sterile
  neutrinos with eV masses in cosmology: How disfavoured exactly?}},
  \href{http://dx.doi.org/10.1088/1475-7516/2011/09/034}{\emph{JCAP} {\bf 1109}
  (2011) 034}, [\href{https://arxiv.org/abs/1108.4136}{{\tt 1108.4136}}].

\bibitem{Giusarma:2011ex}
E.~Giusarma, M.~Corsi, M.~Archidiacono, R.~de~Putter, A.~Melchiorri, O.~Mena
  et~al., \emph{{Constraints on massive sterile neutrino species from current
  and future cosmological data}},
  \href{http://dx.doi.org/10.1103/PhysRevD.83.115023}{\emph{Phys. Rev.} {\bf
  D83} (2011) 115023}, [\href{https://arxiv.org/abs/1102.4774}{{\tt
  1102.4774}}].

\bibitem{Hagstotz:2020ukm}
S.~Hagstotz, P.~F. de~Salas, S.~Gariazzo, M.~Gerbino, M.~Lattanzi, S.~Vagnozzi
  et~al., \emph{{Bounds on light sterile neutrino mass and mixing from
  cosmology and laboratory searches}},
  \href{https://arxiv.org/abs/2003.02289}{{\tt 2003.02289}}.

\bibitem{Feng:2019jqa}
L.~Feng, D.-Z. He, H.-L. Li, J.-F. Zhang and X.~Zhang, \emph{{Constraints on
  active and sterile neutrinos in an interacting dark energy cosmology}},
  \href{https://arxiv.org/abs/1910.03872}{{\tt 1910.03872}}.

\bibitem{Vegetti:2018dly}
S.~Vegetti, G.~Despali, M.~R. Lovell and W.~Enzi, \emph{{Constraining sterile
  neutrino cosmologies with strong gravitational lensing observations at
  redshift $z \sim 0.2$}},
  \href{http://dx.doi.org/10.1093/mnras/sty2393}{\emph{Mon. Not. Roy. Astron.
  Soc.} {\bf 481} (2018) 3661--3669},
  [\href{https://arxiv.org/abs/1801.01505}{{\tt 1801.01505}}].

\bibitem{Zhao:2017urm}
M.-M. Zhao, D.-Z. He, J.-F. Zhang and X.~Zhang, \emph{{Search for sterile
  neutrinos in holographic dark energy cosmology: Reconciling Planck
  observation with the local measurement of the Hubble constant}},
  \href{http://dx.doi.org/10.1103/PhysRevD.96.043520}{\emph{Phys. Rev.} {\bf
  D96} (2017) 043520}, [\href{https://arxiv.org/abs/1703.08456}{{\tt
  1703.08456}}].

\bibitem{Tang:2014yla}
Y.~Tang, \emph{{More Is Different: Reconciling eV Sterile Neutrinos with
  Cosmological Mass Bounds}},
  \href{http://dx.doi.org/10.1016/j.physletb.2015.09.018}{\emph{Phys. Lett.}
  {\bf B750} (2015) 201--208}, [\href{https://arxiv.org/abs/1501.00059}{{\tt
  1501.00059}}].

\bibitem{Vincent:2014rja}
A.~C. Vincent, E.~F. Martinez, P.~Hernández, M.~Lattanzi and O.~Mena,
  \emph{{Revisiting cosmological bounds on sterile neutrinos}},
  \href{http://dx.doi.org/10.1088/1475-7516/2015/04/006}{\emph{JCAP} {\bf 1504}
  (2015) 006}, [\href{https://arxiv.org/abs/1408.1956}{{\tt 1408.1956}}].

\bibitem{Archidiacono:2013xxa}
M.~Archidiacono, N.~Fornengo, C.~Giunti, S.~Hannestad and A.~Melchiorri,
  \emph{{Sterile neutrinos: Cosmology versus short-baseline experiments}},
  \href{http://dx.doi.org/10.1103/PhysRevD.87.125034}{\emph{Phys. Rev.} {\bf
  D87} (2013) 125034}, [\href{https://arxiv.org/abs/1302.6720}{{\tt
  1302.6720}}].

\bibitem{Joudaki:2012uk}
S.~Joudaki, K.~N. Abazajian and M.~Kaplinghat, \emph{{Are Light Sterile
  Neutrinos Preferred or Disfavored by Cosmology?}},
  \href{http://dx.doi.org/10.1103/PhysRevD.87.065003}{\emph{Phys. Rev.} {\bf
  D87} (2013) 065003}, [\href{https://arxiv.org/abs/1208.4354}{{\tt
  1208.4354}}].

\bibitem{Archidiacono:2012ri}
M.~Archidiacono, N.~Fornengo, C.~Giunti and A.~Melchiorri, \emph{{Testing 3+1
  and 3+2 neutrino mass models with cosmology and short baseline experiments}},
  \href{http://dx.doi.org/10.1103/PhysRevD.86.065028}{\emph{Phys. Rev.} {\bf
  D86} (2012) 065028}, [\href{https://arxiv.org/abs/1207.6515}{{\tt
  1207.6515}}].

\bibitem{Archidiacono:2014nda}
M.~Archidiacono, S.~Hannestad, R.~S. Hansen and T.~Tram, \emph{{Cosmology with
  self-interacting sterile neutrinos and dark matter - A pseudoscalar model}},
  \href{http://dx.doi.org/10.1103/PhysRevD.91.065021}{\emph{Phys. Rev.} {\bf
  D91} (2015) 065021}, [\href{https://arxiv.org/abs/1404.5915}{{\tt
  1404.5915}}].

\bibitem{Archidiacono:2015oma}
M.~Archidiacono, S.~Hannestad, R.~S. Hansen and T.~Tram, \emph{{Sterile
  neutrinos with pseudoscalar self-interactions and cosmology}},
  \href{http://dx.doi.org/10.1103/PhysRevD.93.045004}{\emph{Phys. Rev.} {\bf
  D93} (2016) 045004}, [\href{https://arxiv.org/abs/1508.02504}{{\tt
  1508.02504}}].

\bibitem{Archidiacono:2016kkh}
M.~Archidiacono, S.~Gariazzo, C.~Giunti, S.~Hannestad, R.~Hansen, M.~Laveder
  et~al., \emph{{Pseudoscalar-sterile neutrino interactions: reconciling the
  cosmos with neutrino oscillations}},
  \href{http://dx.doi.org/10.1088/1475-7516/2016/08/067}{\emph{JCAP} {\bf 08}
  (2016) 067}, [\href{https://arxiv.org/abs/1606.07673}{{\tt 1606.07673}}].

\bibitem{Hannestad:2013ana}
S.~Hannestad, R.~S. Hansen and T.~Tram, \emph{{How Self-Interactions can
  Reconcile Sterile Neutrinos with Cosmology}},
  \href{http://dx.doi.org/10.1103/PhysRevLett.112.031802}{\emph{Phys. Rev.
  Lett.} {\bf 112} (2014) 031802}, [\href{https://arxiv.org/abs/1310.5926}{{\tt
  1310.5926}}].

\bibitem{Dasgupta:2013zpn}
B.~Dasgupta and J.~Kopp, \emph{{Cosmologically Safe eV-Scale Sterile Neutrinos
  and Improved Dark Matter Structure}},
  \href{http://dx.doi.org/10.1103/PhysRevLett.112.031803}{\emph{Phys. Rev.
  Lett.} {\bf 112} (2014) 031803}, [\href{https://arxiv.org/abs/1310.6337}{{\tt
  1310.6337}}].

\bibitem{Archidiacono:2013dua}
M.~Archidiacono and S.~Hannestad, \emph{{Updated constraints on non-standard
  neutrino interactions from Planck}},
  \href{http://dx.doi.org/10.1088/1475-7516/2014/07/046}{\emph{JCAP} {\bf 1407}
  (2014) 046}, [\href{https://arxiv.org/abs/1311.3873}{{\tt 1311.3873}}].

\bibitem{Forastieri:2019cuf}
F.~Forastieri, M.~Lattanzi and P.~Natoli, \emph{{Cosmological constraints on
  neutrino self-interactions with a light mediator}},
  \href{http://dx.doi.org/10.1103/PhysRevD.100.103526}{\emph{Phys. Rev.} {\bf
  D100} (2019) 103526}, [\href{https://arxiv.org/abs/1904.07810}{{\tt
  1904.07810}}].

\bibitem{Kreisch:2019yzn}
C.~D. Kreisch, F.-Y. Cyr-Racine and O.~Dor{\'e}, \emph{{The Neutrino Puzzle:
  Anomalies, Interactions, and Cosmological Tensions}},
  \href{http://dx.doi.org/10.1103/PhysRevD.101.123505}{\emph{Phys. Rev.} {\bf
  D101} (2020) 123505}, [\href{https://arxiv.org/abs/1902.00534}{{\tt
  1902.00534}}].

\bibitem{Park:2019ibn}
M.~Park, C.~D. Kreisch, J.~Dunkley, B.~Hadzhiyska and F.-Y. Cyr-Racine,
  \emph{{$\Lambda$CDM or self-interacting neutrinos: How CMB data can tell the
  two models apart}},
  \href{http://dx.doi.org/10.1103/PhysRevD.100.063524}{\emph{Phys. Rev.} {\bf
  D100} (2019) 063524}, [\href{https://arxiv.org/abs/1904.02625}{{\tt
  1904.02625}}].

\bibitem{Escudero:2019gvw}
M.~Escudero and S.~J. Witte, \emph{{A CMB Search for the Neutrino Mass
  Mechanism and its Relation to the $H_0$ Tension}},
  \href{http://dx.doi.org/10.1140/epjc/s10052-020-7854-5}{\emph{Eur. Phys. J.
  C} {\bf 80} (2020) 294}, [\href{https://arxiv.org/abs/1909.04044}{{\tt
  1909.04044}}].

\bibitem{Berbig:2020wve}
M.~Berbig, S.~Jana and A.~Trautner, \emph{{The Hubble tension and a
  renormalizable model of gauged neutrino self-interactions}},
  \href{https://arxiv.org/abs/2004.13039}{{\tt 2004.13039}}.

\bibitem{Blinov:2020hmc}
N.~Blinov and G.~Marques-Tavares, \emph{{Interacting radiation after Planck and
  its implications for the Hubble Tension}},
  \href{https://arxiv.org/abs/2003.08387}{{\tt 2003.08387}}.

\bibitem{Mirizzi:2014ama}
A.~Mirizzi, G.~Mangano, O.~Pisanti and N.~Saviano, \emph{{Collisional
  production of sterile neutrinos via secret interactions and cosmological
  implications}},
  \href{http://dx.doi.org/10.1103/PhysRevD.91.025019}{\emph{Phys. Rev.} {\bf
  D91} (2015) 025019}, [\href{https://arxiv.org/abs/1410.1385}{{\tt
  1410.1385}}].

\bibitem{Saviano:2014esa}
N.~Saviano, O.~Pisanti, G.~Mangano and A.~Mirizzi, \emph{{Unveiling secret
  interactions among sterile neutrinos with big-bang nucleosynthesis}},
  \href{http://dx.doi.org/10.1103/PhysRevD.90.113009}{\emph{Phys. Rev.} {\bf
  D90} (2014) 113009}, [\href{https://arxiv.org/abs/1409.1680}{{\tt
  1409.1680}}].

\bibitem{Chu:2015ipa}
X.~Chu, B.~Dasgupta and J.~Kopp, \emph{{Sterile neutrinos with secret
  interactions—lasting friendship with cosmology}},
  \href{http://dx.doi.org/10.1088/1475-7516/2015/10/011}{\emph{JCAP} {\bf 1510}
  (2015) 011}, [\href{https://arxiv.org/abs/1505.02795}{{\tt 1505.02795}}].

\bibitem{Song:2018zyl}
N.~Song, M.~C. Gonzalez-Garcia and J.~Salvado, \emph{{Cosmological constraints
  with self-interacting sterile neutrinos}},
  \href{http://dx.doi.org/10.1088/1475-7516/2018/10/055}{\emph{JCAP} {\bf 1810}
  (2018) 055}, [\href{https://arxiv.org/abs/1805.08218}{{\tt 1805.08218}}].

\bibitem{Chu:2018gxk}
X.~Chu, B.~Dasgupta, M.~Dentler, J.~Kopp and N.~Saviano, \emph{{Sterile
  neutrinos with secret interactions—cosmological discord?}},
  \href{http://dx.doi.org/10.1088/1475-7516/2018/11/049}{\emph{JCAP} {\bf 1811}
  (2018) 049}, [\href{https://arxiv.org/abs/1806.10629}{{\tt 1806.10629}}].

\bibitem{Grohs:2020xxd}
E.~Grohs, G.~M. Fuller and M.~Sen, \emph{{Consequences of neutrino self
  interactions for weak decoupling and big bang nucleosynthesis}},
  \href{https://arxiv.org/abs/2002.08557}{{\tt 2002.08557}}.

\bibitem{Arcadi:2018xdd}
G.~Arcadi, J.~Heeck, F.~Heizmann, S.~Mertens, F.~S. Queiroz, W.~Rodejohann
  et~al., \emph{{Tritium beta decay with additional emission of new light
  bosons}}, \href{http://dx.doi.org/10.1007/JHEP01(2019)206}{\emph{JHEP} {\bf
  01} (2019) 206}, [\href{https://arxiv.org/abs/1811.03530}{{\tt 1811.03530}}].

\bibitem{Blinov:2019gcj}
N.~Blinov, K.~J. Kelly, G.~Z. Krnjaic and S.~D. McDermott, \emph{{Constraining
  the Self-Interacting Neutrino Interpretation of the Hubble Tension}},
  \href{http://dx.doi.org/10.1103/PhysRevLett.123.191102}{\emph{Phys. Rev.
  Lett.} {\bf 123} (2019) 191102},
  [\href{https://arxiv.org/abs/1905.02727}{{\tt 1905.02727}}].

\bibitem{Gariazzo:2019gyi}
S.~Gariazzo, P.~F. de~Salas and S.~Pastor, \emph{{Thermalisation of sterile
  neutrinos in the early Universe in the 3+1 scheme with full mixing matrix}},
  \href{http://dx.doi.org/10.1088/1475-7516/2019/07/014}{\emph{JCAP} {\bf 07}
  (2019) 014}, [\href{https://arxiv.org/abs/1905.11290}{{\tt 1905.11290}}].

\bibitem{Beacom:2004yd}
J.~F. Beacom, N.~F. Bell and S.~Dodelson, \emph{{Neutrinoless universe}},
  \href{http://dx.doi.org/10.1103/PhysRevLett.93.121302}{\emph{Phys. Rev.
  Lett.} {\bf 93} (2004) 121302},
  [\href{https://arxiv.org/abs/astro-ph/0404585}{{\tt astro-ph/0404585}}].

\bibitem{Mangano:2005cc}
G.~Mangano, G.~Miele, S.~Pastor, T.~Pinto, O.~Pisanti and P.~D. Serpico,
  \emph{{Relic neutrino decoupling including flavor oscillations}},
  \href{http://dx.doi.org/10.1016/j.nuclphysb.2005.09.041}{\emph{Nucl.Phys.}
  {\bf B729} (2005) 221--234},
  [\href{https://arxiv.org/abs/hep-ph/0506164}{{\tt hep-ph/0506164}}].

\bibitem{deSalas:2016ztq}
P.~F. de~Salas and S.~Pastor, \emph{{Relic neutrino decoupling with flavour
  oscillations revisited}},
  \href{http://dx.doi.org/10.1088/1475-7516/2016/07/051}{\emph{JCAP} {\bf 07}
  (2016) 051}, [\href{https://arxiv.org/abs/1606.06986}{{\tt 1606.06986}}].

\bibitem{Bennett:2019ewm}
J.~J. Bennett, G.~Buldgen, M.~Drewes and Y.~Y. Wong, \emph{{Towards a precision
  calculation of the effective number of neutrinos $N_{\rm eff}$ in the
  Standard Model I: The QED equation of state}},
  \href{http://dx.doi.org/10.1088/1475-7516/2020/03/003}{\emph{JCAP} {\bf 2003}
  (2020) 003}, [\href{https://arxiv.org/abs/1911.04504}{{\tt 1911.04504}}].

\bibitem{Akita:2020szl}
K.~Akita and M.~Yamaguchi, \emph{{A precision calculation of relic neutrino
  decoupling}},  \href{https://arxiv.org/abs/2005.07047}{{\tt 2005.07047}}.

\bibitem{Schoneberg:2019wmt}
N.~Sch\"oneberg, J.~Lesgourgues and D.~C. Hooper, \emph{{The BAO+BBN take on
  the Hubble tension}},
  \href{http://dx.doi.org/10.1088/1475-7516/2019/10/029}{\emph{JCAP} {\bf 10}
  (2019) 029}, [\href{https://arxiv.org/abs/1907.11594}{{\tt 1907.11594}}].

\bibitem{Lesgourgues:2011re}
J.~Lesgourgues, \emph{{The Cosmic Linear Anisotropy Solving System (CLASS) I:
  Overview}},  \href{https://arxiv.org/abs/1104.2932}{{\tt 1104.2932}}.

\bibitem{Blas:2011rf}
D.~Blas, J.~Lesgourgues and T.~Tram, \emph{{The Cosmic Linear Anisotropy
  Solving System (CLASS) II: Approximation schemes}},
  \href{http://dx.doi.org/10.1088/1475-7516/2011/07/034}{\emph{JCAP} {\bf 1107}
  (2011) 034}, [\href{https://arxiv.org/abs/1104.2933}{{\tt 1104.2933}}].

\bibitem{Lesgourgues:2011rh}
J.~Lesgourgues and T.~Tram, \emph{{The Cosmic Linear Anisotropy Solving System
  (CLASS) IV: efficient implementation of non-cold relics}},
  \href{http://dx.doi.org/10.1088/1475-7516/2011/09/032}{\emph{JCAP} {\bf 1109}
  (2011) 032}, [\href{https://arxiv.org/abs/1104.2935}{{\tt 1104.2935}}].

\bibitem{Audren:2012wb}
B.~Audren, J.~Lesgourgues, K.~Benabed and S.~Prunet, \emph{{Conservative
  Constraints on Early Cosmology: an illustration of the Monte Python
  cosmological parameter inference code}},
  \href{http://dx.doi.org/10.1088/1475-7516/2013/02/001}{\emph{JCAP} {\bf 02}
  (2013) 001}, [\href{https://arxiv.org/abs/1210.7183}{{\tt 1210.7183}}].

\bibitem{Brinckmann:2018cvx}
T.~Brinckmann and J.~Lesgourgues, \emph{{MontePython 3: boosted MCMC sampler
  and other features}},
  \href{http://dx.doi.org/10.1016/j.dark.2018.100260}{\emph{Phys.Dark Univ.}
  {\bf 24} (2019) 100260}, [\href{https://arxiv.org/abs/1804.07261}{{\tt
  1804.07261}}].

\bibitem{Akrami:2018vks}
{\scshape Planck} collaboration, Y.~Akrami et~al., \emph{{Planck 2018 results.
  I. Overview and the cosmological legacy of Planck}},
  \href{https://arxiv.org/abs/1807.06205}{{\tt 1807.06205}}.

\bibitem{Aghanim:2019ame}
{\scshape Planck} collaboration, N.~Aghanim et~al., \emph{{Planck 2018 results.
  V. CMB power spectra and likelihoods}},
  \href{https://arxiv.org/abs/1907.12875}{{\tt 1907.12875}}.

\bibitem{Alam:2016hwk}
{\scshape BOSS} collaboration, S.~Alam et~al., \emph{{The clustering of
  galaxies in the completed SDSS-III Baryon Oscillation Spectroscopic Survey:
  cosmological analysis of the DR12 galaxy sample}},
  \href{http://dx.doi.org/10.1093/mnras/stx721}{\emph{Mon. Not. Roy. Astron.
  Soc.} {\bf 470} (2017) 2617--2652},
  [\href{https://arxiv.org/abs/1607.03155}{{\tt 1607.03155}}].

\bibitem{Beutler:2011hx}
F.~Beutler, C.~Blake, M.~Colless, D.~Jones, L.~Staveley-Smith, L.~Campbell
  et~al., \emph{{The 6dF Galaxy Survey: Baryon Acoustic Oscillations and the
  Local Hubble Constant}},
  \href{http://dx.doi.org/10.1111/j.1365-2966.2011.19250.x}{\emph{Mon. Not.
  Roy. Astron. Soc.} {\bf 416} (2011) 3017--3032},
  [\href{https://arxiv.org/abs/1106.3366}{{\tt 1106.3366}}].

\bibitem{Ross:2014qpa}
A.~J. Ross, L.~Samushia, C.~Howlett, W.~J. Percival, A.~Burden and M.~Manera,
  \emph{{The clustering of the SDSS DR7 main Galaxy sample \textendash{} I. A 4
  per cent distance measure at $z = 0.15$}},
  \href{http://dx.doi.org/10.1093/mnras/stv154}{\emph{Mon. Not. Roy. Astron.
  Soc.} {\bf 449} (2015) 835--847},
  [\href{https://arxiv.org/abs/1409.3242}{{\tt 1409.3242}}].

\bibitem{Bashinsky:2003tk}
S.~Bashinsky and U.~Seljak, \emph{{Neutrino perturbations in CMB anisotropy and
  matter clustering}},
  \href{http://dx.doi.org/10.1103/PhysRevD.69.083002}{\emph{Phys. Rev.} {\bf
  D69} (2004) 083002}, [\href{https://arxiv.org/abs/astro-ph/0310198}{{\tt
  astro-ph/0310198}}].

\bibitem{Kazin:2014qga}
E.~A. Kazin et~al., \emph{{The WiggleZ Dark Energy Survey: improved distance
  measurements to z = 1 with reconstruction of the baryonic acoustic feature}},
  \href{http://dx.doi.org/10.1093/mnras/stu778}{\emph{Mon. Not. Roy. Astron.
  Soc.} {\bf 441} (2014) 3524--3542},
  [\href{https://arxiv.org/abs/1401.0358}{{\tt 1401.0358}}].

\bibitem{Bautista:2017wwp}
J.~E. Bautista et~al., \emph{{The SDSS-IV extended Baryon Oscillation
  Spectroscopic Survey: Baryon Acoustic Oscillations at redshift of 0.72 with
  the DR14 Luminous Red Galaxy Sample}},
  \href{http://dx.doi.org/10.3847/1538-4357/aacea5}{\emph{Astrophys. J.} {\bf
  863} (2018) 110}, [\href{https://arxiv.org/abs/1712.08064}{{\tt
  1712.08064}}].

\bibitem{Blomqvist:2019rah}
M.~Blomqvist et~al., \emph{{Baryon acoustic oscillations from the
  cross-correlation of Ly$\alpha$ absorption and quasars in eBOSS DR14}},
  \href{http://dx.doi.org/10.1051/0004-6361/201935641}{\emph{Astron.
  Astrophys.} {\bf 629} (2019) A86},
  [\href{https://arxiv.org/abs/1904.03430}{{\tt 1904.03430}}].

\bibitem{Pan:2016zla}
Z.~Pan, L.~Knox, B.~Mulroe and A.~Narimani, \emph{{Cosmic Microwave Background
  Acoustic Peak Locations}},
  \href{http://dx.doi.org/10.1093/mnras/stw833}{\emph{Mon. Not. Roy. Astron.
  Soc.} {\bf 459} (2016) 2513--2524},
  [\href{https://arxiv.org/abs/1603.03091}{{\tt 1603.03091}}].

\bibitem{Baumann:2015rya}
D.~Baumann, D.~Green, J.~Meyers and B.~Wallisch, \emph{{Phases of New Physics
  in the CMB}},
  \href{http://dx.doi.org/10.1088/1475-7516/2016/01/007}{\emph{JCAP} {\bf 01}
  (2016) 007}, [\href{https://arxiv.org/abs/1508.06342}{{\tt 1508.06342}}].

\bibitem{Hou:2011ec}
Z.~Hou, R.~Keisler, L.~Knox, M.~Millea and C.~Reichardt, \emph{{How Massless
  Neutrinos Affect the Cosmic Microwave Background Damping Tail}},
  \href{http://dx.doi.org/10.1103/PhysRevD.87.083008}{\emph{Phys. Rev.} {\bf
  D87} (2013) 083008}, [\href{https://arxiv.org/abs/1104.2333}{{\tt
  1104.2333}}].

\bibitem{Giunti:2019aiy}
C.~Giunti and T.~Lasserre, \emph{{eV-scale Sterile Neutrinos}},
  \href{http://dx.doi.org/10.1146/annurev-nucl-101918-023755}{\emph{Ann. Rev.
  Nucl. Part. Sci.} {\bf 69} (2019) 163--190},
  [\href{https://arxiv.org/abs/1901.08330}{{\tt 1901.08330}}].

\bibitem{Diaz:2019fwt}
A.~Diaz, C.~Arg{\"u}elles, G.~Collin, J.~Conrad and M.~Shaevitz, \emph{{Where
  Are We With Light Sterile Neutrinos?}},
  \href{https://arxiv.org/abs/1906.00045}{{\tt 1906.00045}}.

\bibitem{Boser:2019rta}
S.~B{\"o}ser, C.~Buck, C.~Giunti, J.~Lesgourgues, L.~Ludhova, S.~Mertens
  et~al., \emph{{Status of Light Sterile Neutrino Searches}},
  \href{http://dx.doi.org/10.1016/j.ppnp.2019.103736}{\emph{Prog. Part. Nucl.
  Phys.} {\bf 111} (2020) 103736},
  [\href{https://arxiv.org/abs/1906.01739}{{\tt 1906.01739}}].

\bibitem{Aguilar:2001ty}
{\scshape LSND} collaboration, A.~Aguilar-Arevalo et~al., \emph{{Evidence for
  neutrino oscillations from the observation of $\bar{\nu}_e$ appearance in a
  $\bar{\nu}_\mu$ beam}},
  \href{http://dx.doi.org/10.1103/PhysRevD.64.112007}{\emph{Phys. Rev. D} {\bf
  64} (2001) 112007}, [\href{https://arxiv.org/abs/hep-ex/0104049}{{\tt
  hep-ex/0104049}}].

\bibitem{Aguilar-Arevalo:2018gpe}
{\scshape MiniBooNE} collaboration, A.~Aguilar-Arevalo et~al.,
  \emph{{Significant Excess of ElectronLike Events in the MiniBooNE
  Short-Baseline Neutrino Experiment}},
  \href{http://dx.doi.org/10.1103/PhysRevLett.121.221801}{\emph{Phys. Rev.
  Lett.} {\bf 121} (2018) 221801},
  [\href{https://arxiv.org/abs/1805.12028}{{\tt 1805.12028}}].

\bibitem{Laveder:2007zz}
M.~Laveder, \emph{{Unbound neutrino roadmaps}},
  \href{http://dx.doi.org/10.1016/j.nuclphysbps.2007.02.037}{\emph{Nucl. Phys.
  B Proc. Suppl.} {\bf 168} (2007) 344--346}.

\bibitem{Giunti:2006bj}
C.~Giunti and M.~Laveder, \emph{{Short-Baseline Active-Sterile Neutrino
  Oscillations?}},
  \href{http://dx.doi.org/10.1142/S0217732307025455}{\emph{Mod. Phys. Lett. A}
  {\bf 22} (2007) 2499--2509},
  [\href{https://arxiv.org/abs/hep-ph/0610352}{{\tt hep-ph/0610352}}].

\bibitem{Kaether:2010ag}
F.~Kaether, W.~Hampel, G.~Heusser, J.~Kiko and T.~Kirsten, \emph{{Reanalysis of
  the GALLEX solar neutrino flux and source experiments}},
  \href{http://dx.doi.org/10.1016/j.physletb.2010.01.030}{\emph{Phys. Lett. B}
  {\bf 685} (2010) 47--54}, [\href{https://arxiv.org/abs/1001.2731}{{\tt
  1001.2731}}].

\bibitem{Abdurashitov:2005tb}
J.~Abdurashitov et~al., \emph{{Measurement of the response of a Ga solar
  neutrino experiment to neutrinos from an Ar-37 source}},
  \href{http://dx.doi.org/10.1103/PhysRevC.73.045805}{\emph{Phys. Rev. C} {\bf
  73} (2006) 045805}, [\href{https://arxiv.org/abs/nucl-ex/0512041}{{\tt
  nucl-ex/0512041}}].

\bibitem{Mention:2011rk}
G.~Mention, M.~Fechner, T.~Lasserre, T.~Mueller, D.~Lhuillier, M.~Cribier
  et~al., \emph{{The Reactor Antineutrino Anomaly}},
  \href{http://dx.doi.org/10.1103/PhysRevD.83.073006}{\emph{Phys. Rev. D} {\bf
  83} (2011) 073006}, [\href{https://arxiv.org/abs/1101.2755}{{\tt
  1101.2755}}].

\bibitem{Mueller:2011nm}
T.~Mueller et~al., \emph{{Improved Predictions of Reactor Antineutrino
  Spectra}}, \href{http://dx.doi.org/10.1103/PhysRevC.83.054615}{\emph{Phys.
  Rev. C} {\bf 83} (2011) 054615}, [\href{https://arxiv.org/abs/1101.2663}{{\tt
  1101.2663}}].

\bibitem{Huber:2011wv}
P.~Huber, \emph{{On the determination of anti-neutrino spectra from nuclear
  reactors}}, \href{http://dx.doi.org/10.1103/PhysRevC.85.029901}{\emph{Phys.
  Rev. C} {\bf 84} (2011) 024617}, [\href{https://arxiv.org/abs/1106.0687}{{\tt
  1106.0687}}].

\bibitem{Gariazzo:2018mwd}
S.~Gariazzo, C.~Giunti, M.~Laveder and Y.~Li, \emph{{Model-independent
  $\bar\nu_{e}$ short-baseline oscillations from reactor spectral ratios}},
  \href{http://dx.doi.org/10.1016/j.physletb.2018.04.057}{\emph{Phys. Lett. B}
  {\bf 782} (2018) 13--21}, [\href{https://arxiv.org/abs/1801.06467}{{\tt
  1801.06467}}].

\bibitem{Dentler:2018sju}
M.~Dentler, {\'A}.~Hern{\'a}ndez-Cabezudo, J.~Kopp, P.~A. Machado, M.~Maltoni,
  I.~Martinez-Soler et~al., \emph{{Updated Global Analysis of Neutrino
  Oscillations in the Presence of eV-Scale Sterile Neutrinos}},
  \href{http://dx.doi.org/10.1007/JHEP08(2018)010}{\emph{JHEP} {\bf 08} (2018)
  010}, [\href{https://arxiv.org/abs/1803.10661}{{\tt 1803.10661}}].

\bibitem{Declais:1994su}
Y.~Declais et~al., \emph{{Search for neutrino oscillations at 15-meters,
  40-meters, and 95-meters from a nuclear power reactor at Bugey}},
  \href{http://dx.doi.org/10.1016/0550-3213(94)00513-E}{\emph{Nucl. Phys. B}
  {\bf 434} (1995) 503--534}.

\bibitem{Ko:2016owz}
{\scshape NEOS} collaboration, Y.~Ko et~al., \emph{{Sterile Neutrino Search at
  the NEOS Experiment}},
  \href{http://dx.doi.org/10.1103/PhysRevLett.118.121802}{\emph{Phys. Rev.
  Lett.} {\bf 118} (2017) 121802},
  [\href{https://arxiv.org/abs/1610.05134}{{\tt 1610.05134}}].

\bibitem{Danilov:2019aef}
{\scshape DANSS} collaboration, M.~Danilov, \emph{{Recent results of the DANSS
  experiment}},  in \emph{{2019 European Physical Society Conference on High
  Energy Physics}}, 11, 2019.
\newblock \href{https://arxiv.org/abs/1911.10140}{{\tt 1911.10140}}.

\bibitem{Ashenfelter:2018iov}
{\scshape PROSPECT} collaboration, J.~Ashenfelter et~al., \emph{{First search
  for short-baseline neutrino oscillations at HFIR with PROSPECT}},
  \href{http://dx.doi.org/10.1103/PhysRevLett.121.251802}{\emph{Phys. Rev.
  Lett.} {\bf 121} (2018) 251802},
  [\href{https://arxiv.org/abs/1806.02784}{{\tt 1806.02784}}].

\bibitem{Armbruster:1998uk}
B.~Armbruster et~al., \emph{{KARMEN limits on electron-neutrino --->
  tau-neutrino oscillations in two neutrino and three neutrino mixing
  schemes}}, \href{http://dx.doi.org/10.1103/PhysRevC.57.3414}{\emph{Phys. Rev.
  C} {\bf 57} (1998) 3414--3424},
  [\href{https://arxiv.org/abs/hep-ex/9801007}{{\tt hep-ex/9801007}}].

\bibitem{Auerbach:2001hz}
{\scshape LSND} collaboration, L.~Auerbach et~al., \emph{{Measurements of
  charged current reactions of nu(e) on 12-C}},
  \href{http://dx.doi.org/10.1103/PhysRevC.64.065501}{\emph{Phys. Rev. C} {\bf
  64} (2001) 065501}, [\href{https://arxiv.org/abs/hep-ex/0105068}{{\tt
  hep-ex/0105068}}].

\bibitem{Conrad:2011ce}
J.~Conrad and M.~Shaevitz, \emph{{Limits on Electron Neutrino Disappearance
  from the KARMEN and LSND $\nu_e$ - Carbon Cross Section Data}},
  \href{http://dx.doi.org/10.1103/PhysRevD.85.013017}{\emph{Phys. Rev. D} {\bf
  85} (2012) 013017}, [\href{https://arxiv.org/abs/1106.5552}{{\tt
  1106.5552}}].

\bibitem{Giunti:2011cp}
C.~Giunti and M.~Laveder, \emph{{Implications of 3+1 Short-Baseline Neutrino
  Oscillations}},
  \href{http://dx.doi.org/10.1016/j.physletb.2011.11.015}{\emph{Phys. Lett. B}
  {\bf 706} (2011) 200--207}, [\href{https://arxiv.org/abs/1111.1069}{{\tt
  1111.1069}}].

\bibitem{Gariazzo:2017fdh}
S.~Gariazzo, C.~Giunti, M.~Laveder and Y.~Li, \emph{{Updated Global 3+1
  Analysis of Short-BaseLine Neutrino Oscillations}},
  \href{http://dx.doi.org/10.1007/JHEP06(2017)135}{\emph{JHEP} {\bf 06} (2017)
  135}, [\href{https://arxiv.org/abs/1703.00860}{{\tt 1703.00860}}].

\bibitem{Dydak:1983zq}
F.~Dydak et~al., \emph{{A Search for Muon-neutrino Oscillations in the Delta
  m**2 Range 0.3-eV**2 to 90-eV**2}},
  \href{http://dx.doi.org/10.1016/0370-2693(84)90688-9}{\emph{Phys. Lett. B}
  {\bf 134} (1984) 281}.

\bibitem{Stockdale:1984cg}
I.~Stockdale et~al., \emph{{Limits on Muon Neutrino Oscillations in the Mass
  Range 55-eV**2 < Delta m**2 < 800-eV**2}},
  \href{http://dx.doi.org/10.1103/PhysRevLett.52.1384}{\emph{Phys. Rev. Lett.}
  {\bf 52} (1984) 1384}.

\bibitem{Mahn:2011ea}
{\scshape SciBooNE, MiniBooNE} collaboration, K.~Mahn et~al., \emph{{Dual
  baseline search for muon neutrino disappearance at $0.5 {\rm eV}^2 < \Delta
  m^2 < 40 {\rm eV}^2$}},
  \href{http://dx.doi.org/10.1103/PhysRevD.85.032007}{\emph{Phys. Rev. D} {\bf
  85} (2012) 032007}, [\href{https://arxiv.org/abs/1106.5685}{{\tt
  1106.5685}}].

\bibitem{Cheng:2012yy}
{\scshape MiniBooNE, SciBooNE} collaboration, G.~Cheng et~al., \emph{{Dual
  baseline search for muon antineutrino disappearance at $0.1 {\rm eV}^2 <
  {\Delta}m^2 < 100 {\rm eV}^2$}},
  \href{http://dx.doi.org/10.1103/PhysRevD.86.052009}{\emph{Phys. Rev. D} {\bf
  86} (2012) 052009}, [\href{https://arxiv.org/abs/1208.0322}{{\tt
  1208.0322}}].

\bibitem{TheIceCube:2016oqi}
{\scshape IceCube} collaboration, M.~Aartsen et~al., \emph{{Searches for
  Sterile Neutrinos with the IceCube Detector}},
  \href{http://dx.doi.org/10.1103/PhysRevLett.117.071801}{\emph{Phys. Rev.
  Lett.} {\bf 117} (2016) 071801},
  [\href{https://arxiv.org/abs/1605.01990}{{\tt 1605.01990}}].

\bibitem{2005.12942}
M.~Aartsen et~al., \emph{{An eV-scale sterile neutrino search using eight years
  of atmospheric muon neutrino data from the IceCube Neutrino Observatory}},
  \href{https://arxiv.org/abs/2005.12942}{{\tt 2005.12942}}.

\bibitem{2005.12943}
M.~Aartsen et~al., \emph{{Searching for eV-scale sterile neutrinos with eight
  years of atmospheric neutrinos at the IceCube neutrino telescope}},
  \href{https://arxiv.org/abs/2005.12943}{{\tt 2005.12943}}.

\bibitem{Adamson:2017uda}
{\scshape MINOS+} collaboration, P.~Adamson et~al., \emph{{Search for sterile
  neutrinos in MINOS and MINOS+ using a two-detector fit}},
  \href{http://dx.doi.org/10.1103/PhysRevLett.122.091803}{\emph{Phys. Rev.
  Lett.} {\bf 122} (2019) 091803},
  [\href{https://arxiv.org/abs/1710.06488}{{\tt 1710.06488}}].

\bibitem{Maltoni:2007zf}
M.~Maltoni and T.~Schwetz, \emph{{Sterile neutrino oscillations after first
  MiniBooNE results}},
  \href{http://dx.doi.org/10.1103/PhysRevD.76.093005}{\emph{Phys. Rev. D} {\bf
  76} (2007) 093005}, [\href{https://arxiv.org/abs/0705.0107}{{\tt
  0705.0107}}].

\bibitem{Borodovsky:1992pn}
L.~Borodovsky et~al., \emph{{Search for muon-neutrino oscillations
  muon-neutrino ---> electron-neutrino (anti-muon-neutrino --->
  anti-electron-neutrino in a wide band neutrino beam}},
  \href{http://dx.doi.org/10.1103/PhysRevLett.68.274}{\emph{Phys. Rev. Lett.}
  {\bf 68} (1992) 274--277}.

\bibitem{Armbruster:2002mp}
{\scshape KARMEN} collaboration, B.~Armbruster et~al., \emph{{Upper limits for
  neutrino oscillations muon-anti-neutrino ---> electron-anti-neutrino from
  muon decay at rest}},
  \href{http://dx.doi.org/10.1103/PhysRevD.65.112001}{\emph{Phys. Rev. D} {\bf
  65} (2002) 112001}, [\href{https://arxiv.org/abs/hep-ex/0203021}{{\tt
  hep-ex/0203021}}].

\bibitem{Astier:2003gs}
{\scshape NOMAD} collaboration, P.~Astier et~al., \emph{{Search for nu(mu) --->
  nu(e) oscillations in the NOMAD experiment}},
  \href{http://dx.doi.org/10.1016/j.physletb.2003.07.029}{\emph{Phys. Lett. B}
  {\bf 570} (2003) 19--31}, [\href{https://arxiv.org/abs/hep-ex/0306037}{{\tt
  hep-ex/0306037}}].

\bibitem{Antonello:2013gut}
{\scshape ICARUS} collaboration, M.~Antonello et~al., \emph{{Search for
  anomalies in the ${\nu}_e$ appearance from a ${\nu}_{\mu}$ beam}},
  \href{http://dx.doi.org/10.1140/epjc/s10052-013-2599-z}{\emph{Eur. Phys. J.
  C} {\bf 73} (2013) 2599}, [\href{https://arxiv.org/abs/1307.4699}{{\tt
  1307.4699}}].

\bibitem{Agafonova:2013xsk}
{\scshape OPERA} collaboration, N.~Agafonova et~al., \emph{{Search for $\nu_\mu
  \rightarrow \nu_e$ oscillations with the OPERA experiment in the CNGS beam}},
  \href{http://dx.doi.org/10.1007/JHEP07(2013)004}{\emph{JHEP} {\bf 07} (2013)
  004}, [\href{https://arxiv.org/abs/1303.3953}{{\tt 1303.3953}}].

\end{thebibliography}\endgroup

\end{document}